\documentclass[letterpaper,12pt]{article}   
\usepackage{osajnl2} 
\usepackage[draft]{hyperref} 
\usepackage{amssymb}
\usepackage{rotating}
\rotdriver{dvips}
\newcommand{\bc}{\begin{center}}
\newcommand{\ec}{\end{center}}
\def\jqsrt{J.\ Quant.\ Spectrosc.\ Rad.\ Transfer\ }
\def\degr{\hbox{$^\circ$}}
\def\be{\begin{equation}}
\def\ee{\end{equation}}
\newcommand{\ba}{\begin{array}}
\newcommand{\ea}{\end{array}}
\def\bea{\begin{eqnarray}}
\def\eea{\end{eqnarray}}

\def\la{\mathrel{\mathchoice {\vcenter{\offinterlineskip\halign{\hfil
$\displaystyle##$\hfil\cr<\cr\sim\cr}}}
{\vcenter{\offinterlineskip\halign{\hfil$\textstyle##$\hfil\cr
<\cr\sim\cr}}}
{\vcenter{\offinterlineskip\halign{\hfil$\scriptstyle##$\hfil\cr
<\cr\sim\cr}}}
{\vcenter{\offinterlineskip\halign{\hfil$\scriptscriptstyle##$\hfil\cr
<\cr\sim\cr}}}}}
\def\ga{\mathrel{\mathchoice {\vcenter{\offinterlineskip\halign{\hfil
$\displaystyle##$\hfil\cr>\cr\sim\cr}}}
{\vcenter{\offinterlineskip\halign{\hfil$\textstyle##$\hfil\cr
>\cr\sim\cr}}}
{\vcenter{\offinterlineskip\halign{\hfil$\scriptstyle##$\hfil\cr
>\cr\sim\cr}}}
{\vcenter{\offinterlineskip\halign{\hfil$\scriptscriptstyle##$\hfil\cr
>\cr\sim\cr}}}}}

\begin{document}

\title{Light scattering by a multilayered spheroidal particle}

\author{Victor G. Farafonov$^{1}$ and Nikolai V. Voshchinnikov$^{2,*}$}

\address{$^1$ State University of Aerospace Instrumentation,
St.~Petersburg, 190000 Russia}
\address{$^2$ Sobolev Astronomical Institute, St.~Petersburg University,
St.~Petersburg, 198504 Russia}
\address{$^*$Corresponding author: nvv@astro.spbu.ru}

\begin{abstract}

The light scattering problem for a
confocal multilayered spheroid has been solved by the
extended boundary condition method (EBCM)
with a corresponding spheroidal basis.
The solution preserves the  advantages of the approach
applied previously to homogeneous and core-mantle spheroids, i.e.
the separation of the  radiation fields into two parts and
a special choice of scalar potentials for each of the parts.
The method is known to be useful in a wide range of the particle parameters.
It is particularly efficient for strongly prolate and oblate spheroids.
Numerical tests are described.
Illustrative calculations have shown that
the extinction factors to converge
to average values with a growing number of layers
and how the extinction vary with a growth of particle porosity.
\end{abstract}


\maketitle 

\section{Introduction}

A detailed knowledge of the optics of inhomogeneous (layered) non-spherical
particles is required in many scientific and industrial applications.
Numerical treatment of these particles is a very complicated problem, especially
when the particle size is not small and (or) the particle shape
appreciably deviates from spherical
(see \cite{kan03, fip03, vfi09} for a review of available methods).
However, for one of the simplest cases,
multilayered spheroids,  rather fast calculations of a high
accuracy can be performed by the method of separation of variables (SVM)
and the extended boundary condition method (EBCM).
Both methods can be used with spherical or spheroidal basis
so that  the electromagnetic
fields are expanded in terms of spherical or spheroidal
wave functions, respectively \cite{vfi09, f11}.
As a result, characteristics of scattered radiation can be calculated by using
the same expressions. Note that the methods differ
in formulation of the boundary conditions
(see \cite{f11} for detailed discussion).
However, the use of the spherical basis
is not appropriate for particles of large eccentricity
(with aspect ratios $a/b \ga 1.5 - 2$) that is why one needs to
apply a spheroidal basis for these particles
when geometry of the problem is sorely taken into account.

The first attempt to develop a solution for multilayered
confocal spheroids by SVM with a spheroidal basis
was made in \cite{getal00} by using a recursive procedure when
passing from one layer to the next.
The paper did not contain calculations because they
require solving a complex nonlinear matrix
equation for the unknown expansion coefficients.
Later the algorithm was modified by using the ideas presented
in \cite{far01} and some numerical results were
published in \cite{getal03} for small particles  with large
refractive indices.
Using the SVM approach, an exact solution for spheroids with
non-confocal layers was obtained but the calculations were
published for core-mantle particles only \cite{han06}.

Layered axisymmetric particles (including spheroids) were also
treated by the $T$-matrix method (e.g., \cite{wb79, pet07}) and
generalized multipole technique or null-field method
(e.g., \cite{dwe06}). However, these methods
did not provide the adequate numerical results
for strongly nonspherical particles with a large number of layers
(see discussion in \cite{vfi09}).

In this paper, we consider the scattering of an arbitrary polarized plane wave by
confocal multilayered spheroids.
We have developed the recursive EBCM solution with a spheroidal basis suggested
in \cite{far01} by taking into account inaccuracies found during our numerical
realization of the algorithm (see also \cite{nvv06}).
Our solution is based on a special choice of scalar potentials which
for any next layer can be found by using the potentials of the previous layer,
the procedure starting from the particle core.
These potentials are expanded in terms of spheroidal wave functions.
The unknown expansion coefficients  of scattered radiation potentials
are determined by solving the systems of linear matrix equations.
It is important to emphasize
that the dimension of these systems for layered spheroids
does not increase as compared to that for homogeneous spheroids
which is contrary to the SVM (see, e.g., \cite{fvs96}).
Calculations show that the method suggested in this paper
gives results of high accuracy
and can be used in a wide range of the particle parameters.
This confirms the conclusion made in \cite{vfi09} that
the EBCM with a corresponding spheroidal basis is more
preferable in the treatment of  multilayered confocal spheroids.

\section{Formulation of the Problem}

The problem of electromagnetic light scattering by a multilayered spheroidal
particle is solved in the prolate and oblate spheroidal coordinate systems
($\xi$,\,$\eta$,\,$\varphi$) which are connected with the Cartesian system
($x,\,y,\,z$) in the following way \cite{F57, KPS76}:
\begin{eqnarray}
x & = & \frac{d}{2} \, ( \xi^{2} - \tilde{f} )^{1/2} (1-\eta^{2})^{1/2}
\cos{\varphi}, \nonumber \\
y & = & \frac{d}{2}\, ( \xi^{2} - \tilde{f} )^{1/2} (1-\eta^{2})^{1/2} \sin{\varphi},\\
z & = & \frac{d}{2}\, \xi \eta, \nonumber
\end{eqnarray}
where $\tilde{f}=1$, $\xi$ $\in$ [1,\,$\infty$),\,\, $\eta$ $\in$ [--1,\,1], $\varphi$ $\in$
[0,\,2$\pi$)  for prolate coordinates and $\tilde{f}=-1$,
$\xi$ $\in$ [0,\,$\infty$), \, $\eta$ $\in$ [--1,\,1], $\varphi$ $\in$
[0,\,2$\pi$) for oblate coordinates; \it{d} \rm is the focal distance.
We assume that the particle is confocal. This means that the
surfaces of layers coincide with the coordinate surfaces, and
their equations can be written as
\be
\xi = \xi_j ,
\ee
where
$j=1, 2, \ldots , N$ ($N \geq 3$ is the number of layers,
$j=1$ for the outermost surface, i.e. the particle boundary, and
$j=N$ for the boundary of the core).
For such particles, the major and minor semiaxes of the shell spheroids,
$a_j$ and $b_j$, satisfy the following conditions:
\be
a_1^2 - b_1^2 = a_2^2 - b_2^2 = \  \cdots \ = a_N^2 - b_N^2 \ =
\ \left( \frac{d}{2} \right)^2 .
\ee

Let the time-dependent part of the electromagnetic field  be
$\exp{(-i \omega t)}$ and
$\vec E$, $\vec H$
be the vectors of the electric and magnetic fields, respectively.
The vectors $\vec E^{(0)}$, $\vec H^{(0)}$ correspond to
the field of the incident radiation,
$\vec E^{(1)}$, $\vec H^{(1)}$  to the field of the scattered radiation,
$\vec E^{(2)}$, $\vec H^{(2)}$  to the field inside the outermost layer,
$\dots$,
$\vec E^{(j)}$, $\vec H^{(j)}$  to the field inside the ($j-1$)th layer,
$\dots$,
$\vec E^{(N+1)}$, $\vec H^{(N+1)}$ to the field the particle core.

We consider  a plane electromagnetic wave with an arbitrary
polarization propagating at an incident angle $\alpha$ to the rotational
axis of the spheroid (or the $z$-axis; see Fig.~\ref{geon}). This wave can be represented
as a superposition of two components
(the magnetic fields can be obtained from the electrical ones by using
Maxwell's equations): \\
(a) TE mode
\be
\vec E^{ (0) }  =  - \vec i_{y} \exp \left [ {i k_{1} (x \sin{\alpha} + z
\cos{\alpha} ) } \right ]; \label{te0}
\ee
(b) TM mode
\be
\vec E^{ (0) }  =  ( \vec i_{x} \cos{ \alpha } - \vec i_{z} \sin {\alpha} )
\exp \left [ {i k_{1} (x \sin{\alpha} + z \cos{\alpha} ) } \right ]. \label{tm0}
\ee
Here $\vec i_{x}, \, \vec i_{y}, \, \vec i_{z}$ are the unit vectors in the
Cartesian coordinate system,
$k_{i} = \sqrt{ \varepsilon_{i} \mu_{i} } k_{0}$ is the wave number in a
medium with the complex permettivity $\varepsilon_{i}$ and the magnetic
permeability $\mu_{i}$,
$k_{0} = 2\pi/\lambda_0$  and $k_{1} = 2\pi/\lambda_1$ are the wave numbers
in vacuum and the medium outside the particle, respectively.

As it has been previously shown in \cite{fi06} (see also \cite{vf93, fvs96}),
in the case of axisymmetric particles, the scattering problem can
be solved independently for each term of the Fourier expansion of the vectors
$\vec E^{(i)}$ and $\vec H^{(i)}$
in terms of the azimuthal angle $\varphi$.
In the following, we represent all electromagnetic fields as
\begin{equation}
\vec E^{(i)} = \vec E^{(i)}_{1} + \vec E^{(i)}_{2}, \,\,\,\,\,\,
\vec H^{(i)} = \vec H^{(i)}_{1} + \vec H^{(i)}_{2}, \,\,\,\,\,\,
i = 0, 1, 2, \ldots, N+1
\label{field}\end{equation}
so that $\vec E^{(i)}_{1}$ and $\vec H^{(i)}_{1}$
are independent of the azimuthal angle $\varphi$
(the zeroth term of the Fourier series), whereas the averaging of
$\vec E^{(i)}_{2}$ and $\vec H^{(i)}_{2}$ over $\varphi$ gives zero.
Below, the axisymmetric problem for
the fields $\vec E^{(i)}_{1}$, $\vec H^{(i)}_{1}$
and the non-axisymmetric problem
for the fields $\vec E^{(i)}_{2}$,  $\vec H^{(i)}_{2}$
are solved independently of one another.

\section{Solution to the Axisymmetric Problem}

Let us consider the scalar potentials
\begin{equation}
{\cal P}^{(i)} = E_{1\varphi}^{(i)} \,\cos\varphi, \,\,\,\,\,\,
{\cal Q}^{(i)} = H_{1\varphi}^{(i)} \,\cos\varphi,
\end{equation}
where $E_{1\varphi}^{(i)}$,  $H_{1\varphi}^{(i)}$ are
the $\varphi$-components of the vectors $\vec E^{(i)}_{1}$, $\vec H^{(i)}_{1}$
($i = 0, 1, 2, \ldots, N+1$).
If we remove the  $\cos\varphi$ factor,
these potentials coincide with the Abraham potentials
within the factor
$h_{\varphi} = ({d}/{2}) \sqrt {(\xi^2 - \tilde{f} )(1 -\eta^2})$
\cite{fi06}.
It follows from Maxwell's equations that the scalar
potentials satisfy the wave (Helmholtz) equations
\be \Delta {\cal P}^{(i)}  + k^2_i \, {\cal P}^{(i)}  =0, \ \ \ \
\Delta  {\cal Q}^{(i)} + k^2_i \,  {\cal Q}^{(i)} =0.
\label{wave}\ee
The remaining components of the electromagnetic
fields $E_{1\varphi}^{(i)}$,  $H_{1\varphi}^{(i)}$
can be expressed in terms of their azimuthal components.
Note that the axisymmetric problem is solved independently
for potentials ${\cal P}$ and ${\cal Q}$, i.e., for the TE and TM waves.

In the case of the TE mode (see Eq.~(\ref{te0})), the boundary conditions
(the continuity of the tangential components of
the electromagnetic fields at the interfaces)
should be rewritten as
\be
\left.
\ba{l}
 {\cal P}^{(0)} + {\cal P}^{(1)} = {\cal P}^{(2)} , \\
\displaystyle \frac{\partial \left[\sqrt{ \xi^2 - \tilde{f}}
({\cal P}^{(0)} + {\cal P}^{(1)}) \right] }{\partial \xi}
= \frac{\mu_1}{\mu_{2}} \,
\frac{\partial \left[ \sqrt{ \xi^2 - \tilde{f}}{\cal P}^{(2)} \right]}
{\partial \xi},
\ea  \right\}_{ \xi = \xi_1}
\label{11}
\ee
\be
\left.
\ba{l}
 {\cal P}^{(j)} = {\cal P}^{(j+1)} , \\
\displaystyle \frac{\partial \left[\sqrt{ \xi^2 - \tilde{f}}{\cal P}^{(j)} \right] }{\partial \xi}
= \frac{\mu_j}{\mu_{j+1}} \,
\frac{\partial \left[ \sqrt{ \xi^2 - \tilde{f}}{\cal P}^{(j+1)} \right] }
{\partial \xi}, \\
\ea \right\}_{ \xi = \xi_j}
\label{12} \ee
where  $j=2, 3, \ldots , N$.

Let us next formulate the problem in the form of surface integral equations.
We can represent the potential ${\cal P}^{(j)}$ of the radiation in the
($j - 1$)th shell of the particle as ($j=2, 3, \ldots , N+1$)
\be
 {\cal P}^{(j)} = {\cal P}^{(j)}_{\rm A} + {\cal P}^{(j)}_{\rm B},
\ee
where ${\cal P}^{(j)}_{\rm A}$ has no singularities in the
region $D_{j-1}$ (and hence in the region $D_j$ enclosed by the surface
$S_j$) and the potential ${\cal P}^{(j)}_{\rm B}$
satisfies the radiation condition at infinity.
Note that inside the core
${\cal P}^{(N+1)} = {\cal P}^{(N+1)}_{\rm A}$,
that is, ${\cal P}^{(N+1)}_{\rm B} = 0$.
Within the framework of EBCM, we obtain the system of surface integral
equations (see \cite{far01} for more details)
\bea & & \frac{d}{2} \, (\xi^2_1 - \tilde{f}) \int_{0}^{2 \pi} \int_{0}^{ \pi}
\left\{ {\cal P}^{(2)} \left( \vec{r}^{\,\prime} \right) \frac{\partial
G_1 }{\partial \xi^\prime} - \left[ \frac{\mu_1}{\mu_{2}} \,
\frac{\partial {\cal P}^{(2)} \left( \vec{r}^{\,\prime} \right)}
{\partial \xi^\prime} + \left( \frac{\mu_1}{\mu_{2}} - 1 \right)
\right. \right. \nonumber
\\ & & \left. \left. \times \frac{\xi_1}{(\xi^2_1 - \tilde{f})} \, {\cal P}^{(2)}
\left( \vec{r}^{\,\prime} \right)  \right] G_1 \right \} {\rm d }
\eta^\prime {\rm d } \varphi^\prime = \left\{ \ba{lcl} -{\cal P}^{(0)}
(\vec{r}), \ \ \  \vec{r} \in { D_1},
\\
\phantom{-}{\cal P}^{(1)} (\vec{r}), \ \ \   \vec{r} \in R^3 \setminus \bar{ D_1},
\ea \right. \label{190}
\eea
where $\xi^\prime = \xi_1$,
\bea & & \frac{d}{2} \, (\xi^2_j - \tilde{f}) \int_{0}^{2 \pi} \int_{0}^{ \pi}
\left\{ {\cal P}^{(j+1)} \left( \vec{r}^{\,\prime} \right) \frac{\partial
G_j }{\partial \xi^\prime} - \left[ \frac{\mu_j}{\mu_{j+1}} \,
\frac{\partial {\cal P}^{(j+1)} \left( \vec{r}^{\,\prime} \right)}
{\partial \xi^\prime} + \left( \frac{\mu_j}{\mu_{j+1}} - 1 \right)
\right. \right. \nonumber
\\ & & \left. \left. \times \frac{\xi_j}{(\xi^2_j - \tilde{f})} \, {\cal P}^{(j+1)}
\left( \vec{r}^{\,\prime} \right)  \right] G_j \right \} {\rm d }
\eta^\prime {\rm d } \varphi^\prime = \left\{ \ba{lcl} -{\cal P}^{(j)}_{\rm A}
(\vec{r}), \ \ \  \vec{r} \in { D_j},
\\
\phantom{-}{\cal P}^{(j)}_{\rm B} (\vec{r}), \ \ \   \vec{r} \in R^3 \setminus \bar{ D_j},
\ea \right. \label{19}
\eea
where $\xi^\prime = \xi_j$,
\be
G_j = G(k_j,\vec{r},\vec{r}^{\,\prime}) = \frac{\exp{ik_j |\vec{r} -
\vec{r}^{\,\prime}|}} {4 \pi |\vec{r} - \vec{r}^{\,\prime}|}
\ee
is the Green function of the wave equation with the wave number $k_j$,
$j= 2, 3, \ldots , N$.

The scalar potentials can be expanded in terms of the
spheroidal functions \cite{KPS76}
\be \ba{l}
{\cal P}^{(0)} \\
{\cal Q}^{(0)}
\ea
= \sum_{l=1}^\infty\
\ba{l}
a^{(0)}_{l} \\
b^{(0)}_{l}
\ea
R_{1l}^{(1)} (c_1, \xi)\ S_{1l} (c_1, \eta)\ \cos\varphi,
\label{eqA0}\ee
\be
\ba{l}
{\cal P}^{(1)} \\
{\cal Q}^{(1)}
\ea
= \sum_{l=1}^\infty\
\ba{l}
a^{(1)}_{l} \\
b^{(1)}_{l} \ea R_{1l}^{(3)} (c_1, \xi)\ S_{1l} (c_1, \eta)\
\cos\varphi,
\label{eqB1}\ee
\be \ba{l}
{\cal P}^{(j)}_{\rm A} \\
{\cal Q}^{(j)}_{\rm A}
\ea
= \sum_{l=1}^\infty\
\ba{l}
a^{(j)}_{l} \\
b^{(j)}_{l}
\ea
R_{1l}^{(1)} (c_j, \xi)\ S_{1l} (c_j, \eta)\ \cos\varphi,
\label{eqA}\ee
\be
\ba{l}
{\cal P}^{(j)}_{\rm B} \\
{\cal Q}^{(j)}_{\rm B}
\ea
= \sum_{l=1}^\infty\
\ba{l}
c^{(j)}_{l} \\
d^{(j)}_{l} \ea R_{1l}^{(3)} (c_j, \xi)\ S_{1l} (c_j, \eta)\
\cos\varphi,
\label{eqB}\ee
where $j= 2, 3, \ldots , N+1$.
For the incident radiation we obtain the following coefficients
\cite{vf93, fvs96}: \\
(a) TE mode (see Eq.~(\ref{te0}))
\begin{equation} \begin{array}{l}
a^{(0)}_{l} = - 2i^{l} N_{1l}^{-2}(c_{1}) S_{1l}^{}(c_{1}, \cos{\alpha}),
\,\,\,  b^{(0)}_{l} = 0;
\end{array} \label{tee0} \end{equation}
(b) TM mode (see Eq.~(\ref{tm0}))
\begin{equation} \begin{array}{l}
a^{(0)}_{l} = 0, \,\,\,\,
\displaystyle b^{(0)}_{l} = 2i^{l}\sqrt{\frac{\varepsilon_{1}}{\mu_{1}}}
N^{-2}_{1l}(c_{1}) S_{1l}(c_{1}, \cos{\alpha}).
\end{array} \end{equation}
Here, $ R_{ml}^{(1),(3)}(c_j, \xi)$ are the prolate radial spheroidal
functions of the first and third kinds,
$S_{ml}(c_{j}, \eta)$ the prolate angular
functions with the normalization coefficients $N_{ml}(c_{j})$ \cite{KPS76},
and the parameter $c_{j}=k_{j}(d/2)$. 

For the expansion of the Green function in terms of the
spheroidal functions, we have
\cite{KPS76}:
\bea G (k_j, \vec{r},
\vec{r}^{\,\prime}) & = & \frac{ i k_{j} }{ 2 \pi }
\sum_{m=0}^\infty \sum_{l=m}^\infty ( 2 - \delta_{0m} )\,
N_{ml}^{-2}(c_j)\, R_{ml}^{(1)} (c_j, \xi_{<})\, R_{ml}^{(3)} (c_j, \xi_{>})\,
\nonumber \\
& &  \times
S_{ml} (c_j, \eta)\, S_{ml} (c_j, \eta^\prime)\, \cos m(\varphi - \varphi^\prime),
\label{28}
\eea
where
$$
\delta_{0m} = \left\{
\ba{l}
1,\  \ m=0, \\
0,\ \  m \neq 0,
\ea \right. \nonumber
$$
and $\xi_{<} = \min (\xi,\xi^\prime),\,\xi_{>} = \max (\xi, \xi^\prime).$

We substitute Eqs.~(\ref{eqA0})--(\ref{eqB}), and (\ref{28}) into
the integral equations (\ref{190}), (\ref{19}).
Taking into account orthogonality of the angular spheroidal
functions $S_{ml}(c_j,\eta) \cos m\varphi$ on the surface of any spheroid,
we obtain the linear algebraic
equations for the unknown expansion coefficients
of the potentials considered. In the matrix notation,
they have the following form ($j=1, 2, \ldots , N$):
\be
 \left( \begin{array}{c}
\vec{z}^{(j)}_{\rm A} \\
\vec{z}^{(j)}_{\rm B}
          \end{array} \right) =
{-{\cal A}^{(j)}_{31}  -{\cal A}^{(j)}_{33}
\choose
{\cal A}^{(j)}_{11} \ \ \ \ \ \  {\cal A}^{(j)}_{13}} \
 \left( \begin{array}{c}
\vec{z}^{(j+1)}_{\rm A} \\
\vec{z}^{(j+1)}_{\rm B}
          \end{array} \right),
 \label{26} \ee
where
\bea
{\cal A}^{(j)}_{31} & = &
{\cal W}_j \ \left[ {\cal R}^{[3]}(c_j, \xi_j)\
\Delta^{(1)}(c_j, c_{j+1})
- \frac{\mu_j}{\mu_{j+1}}
\Delta^{(1)}(c_j, c_{j+1})\ {\cal R}^{[1]}(c_{j+1}, \xi_j)
\right.  \nonumber \\ & & \left.
- \left( \frac{\mu_j}{\mu_{j+1}} - 1 \right)\
\frac{\xi_j}{\xi^2_j - \tilde{f}}\ \Delta^{(1)}(c_j, c_{j+1})
\right] \ {\cal P}^{[1]}(c_{j+1}, \xi_j, \xi_{j+1}),
\label{a0-31}\eea
\bea
{\cal A}^{(j)}_{33} & = &
{\cal W}_j \ \left[ {\cal R}^{[3]}(c_j, \xi_j)\
\Delta^{(1)}(c_j, c_{j+1})
- \frac{\mu_j}{\mu_{j+1}}
\Delta^{(1)}(c_j, c_{j+1})\ {\cal R}^{[3]}(c_{j+1}, \xi_j)
\right.  \nonumber \\ & & \left.
- \left( \frac{\mu_j}{\mu_{j+1}} - 1 \right)\
\frac{\xi_j}{\xi^2_j - \tilde{f}}\ \Delta^{(1)}(c_j, c_{j+1})
\right] \ {\cal P}^{[3]}(c_{j+1}, \xi_j, \xi_{j+1}),
\label{a0-33}\eea
\bea
{\cal A}^{(j)}_{11} & = &
{\cal W}_j \ \left[ {\cal R}^{[1]}(c_j, \xi_j)\
\Delta^{(1)}(c_j, c_{j+1})
- \frac{\mu_j}{\mu_{j+1}}
\Delta^{(1)}(c_j, c_{j+1})\ {\cal R}^{[1]}(c_{j+1}, \xi_j)
\right.  \nonumber \\ & & \left.
- \left( \frac{\mu_j}{\mu_{j+1}} - 1 \right)\
\frac{\xi_j}{\xi^2_j - \tilde{f}}\ \Delta^{(1)}(c_j, c_{j+1})
\right] \ {\cal P}^{[1]}(c_{j+1}, \xi_j, \xi_{j+1}),
\label{a0-11}\eea
\bea
{\cal A}^{(j)}_{13} & = &
{\cal W}_j \ \left[ {\cal R}^{[1]}(c_j, \xi_j)\
\Delta^{(1)}(c_j, c_{j+1})
- \frac{\mu_j}{\mu_{j+1}}
\Delta^{(1)}(c_j, c_{j+1})\ {\cal R}^{[3]}(c_{j+1}, \xi_j)
\right.  \nonumber \\ & & \left.
- \left( \frac{\mu_j}{\mu_{j+1}} - 1 \right)\
\frac{\xi_j}{\xi^2_j - \tilde{f}}\ \Delta^{(1)}(c_j, c_{j+1})
\right] \ {\cal P}^{[3]}(c_{j+1}, \xi_j, \xi_{j+1}).
\label{a0-13}\eea
Above, we introduce the vectors  specified by
\be
 \left( \begin{array}{c}
\vec{z}^{(1)}_{\rm A} \\
\vec{z}^{(1)}_{\rm B}
          \end{array} \right) =
\left( \begin{array}{c}
         \left\{ a^{(0)}_{l} R^{(1)}_{1l}(c_{1}, \xi_{1})\,N_{1l}(c_{1}) \right\}^{\infty}_{1\phantom{_k}} \\
         \left\{ a^{(1)}_{l} R^{(3)}_{1l}(c_{1}, \xi_{1})\,N_{1l}(c_{1}) \right\}^{\infty}_{1}
          \end{array} \right),
\ee
\be
 \left( \begin{array}{c}
\vec{z}^{(j)}_{\rm A} \\
\vec{z}^{(j)}_{\rm B}
          \end{array} \right) =
\left( \begin{array}{c}
         \left\{ a^{(j)}_{l} R^{(1)}_{1l}(c_{j}, \xi_{j})\,N_{1l}(c_{j}) \right\}^{\infty}_{1\phantom{_k}} \\
         \left\{ c^{(j)}_{l} R^{(3)}_{1l}(c_{j}, \xi_{j})\,N_{1l}(c_{j}) \right\}^{\infty}_{1}
          \end{array} \right),
\ee
where $j=2, \ldots , N$ and the diagonal matrices
\be
{\cal R}^{[i]}(c_j, \xi_j) = \left \{
{R^{(i)}_{ml}}^{\prime} (c_{j}, \xi_j) / R^{(i)}_{ml} (c_{j}, \xi_j)
\delta_{nl} \right\}^{\infty}_{m},
\ee
\be
{\cal W}_j = - \left[{\cal R}^{[3]}(c_j, \xi_j) - {\cal R}^{[1]}(c_j, \xi_j) \right]^{-1},
\label{eqw}\ee
\be
{\cal P}^{[i]}(c_j, \xi_{j-1}, \xi_j)
= \left\{ R^{(i)}_{ml}(c_{j}, \xi_{j-1}) / R^{(i)}_{ml}(c_{j}, \xi_{j})
\delta_{nl} \right\}^{\infty}_{m}.
\ee
The matrix elements
$\Delta^{(m)}(c_j, c_{j+1})  = \left\{ \delta^{(m)}_{nl} (c_{j}, c_{j+1})
\right\}^{\infty}_{m\phantom{_k}}$
are integrals of the products of the angular spheroidal functions \cite{fvs96}.
To derive  Eq.~(\ref{eqw}), we use  expression for the Wronskian
of the radial spheroidal functions \cite{KPS76}.
Since $\vec{z}^{(N+1)}_{\rm B}=0$, the
system of equations (\ref{26}) can be easily solved
relative to
the expansion coefficients of the scattered radiation potential
\be
\vec{z}^{(1)}_{\rm B} = {\cal A}_2  {\cal A}_1^{(-1)} \ \vec{z}^{(1)}_{\rm A},
\label{ek}\ee
where
the coefficients of the incident radiation are given by Eq.~(\ref{tee0}).
The matrices ${\cal A}_1$ and ${\cal A}_2$ satisfy the relation
\be
{{\cal A}_1 \choose {\cal A}_2}
=
{-{\cal A}^{(1)}_{31}  -{\cal A}^{(1)}_{33}
\choose
  {\cal A}^{(1)}_{11} \ \ \ \ \ \  {\cal A}^{(1)}_{13}}
\ \cdots \
{-{\cal A}^{(j)}_{31}  -{\cal A}^{(j)}_{33}
\choose
  {\cal A}^{(j)}_{11} \ \ \ \ \ \  {\cal A}^{(j)}_{13}}
\
\cdots \
{-{\cal A}^{(N-1)}_{31}  -{\cal A}^{(N-1)}_{33}
\choose
  {\cal A}^{(N-1)}_{11} \ \ \ \ \ \  {\cal A}^{(N-1)}_{13}}
{-{\cal A}^{(N)}_{31}
\choose
 \ \ {\cal A}^{(N)}_{11} }.
\label{matr}\ee

An important point is that the representation of Eqs.~(\ref{ek}), (\ref{matr})
in the recursive form needs only one matrix inversion
in contrast with the T-matrix representation that requires inversions
for each layer and one more at the last step
(see  discussion in \cite{fip03}).

For TM mode, the transformation of the  above equations
for potentials ${\cal Q}^{(j)}$
is performed by the replacements $\mu_{j} \rightarrow
\varepsilon_{j}, \,\,\, \varepsilon_{j} \rightarrow \mu_{j}$,
 $a^{(j)}_{l}  \rightarrow b^{(j)}_{l}$,
and $c^{(j)}_{l}  \rightarrow  d^{(j)}_{l}$.
In order to obtain the corresponding systems for oblate spheroid one must use
the standard replacements $c \rightarrow - i c$,
$\xi \rightarrow i \xi$ and  oblate
spheroidal functions instead of the prolate ones.
For example,  in the case of oblate spheroids and TM mode, Eq.~(\ref{a0-31})
can be written as:
\bea
{\cal A}^{(j)}_{31} & = &
{\cal W}_j \ \left[ {\cal R}^{[3]}(-ic_j, i \xi_j)\
\Delta^{(1)}(-ic_j, -ic_{j+1})
- \frac{\varepsilon_j}{\varepsilon_{j+1}}
\Delta^{(1)}(-ic_j, -ic_{j+1})\ {\cal R}^{[1]}(-ic_{j+1}, i \xi_j)
\right.  \nonumber \\ & & \left.
- \left( \frac{\varepsilon_j}{\varepsilon_{j+1}} - 1 \right)\
\frac{i \xi_j}{(i \xi)^2_j - \tilde{f}}\ \Delta^{(1)}(-ic_j, -ic_{j+1})
\right] \ {\cal P}^{[1]}(-ic_{j+1}, i \xi_j, i \xi_{j+1}).  \nonumber
\eea

\section{Solution to the Non-Axisymmetric Problem}

The second terms in Eqs.~(\ref{field}) can be represented in the following form: \\
(a) TE mode
\begin{eqnarray}
\vec{E}^{(i)}_{2} & = & \vec \nabla \times \left( {U}^{(i)} \vec{i}_{z} + {V}^{(i)}
\vec{r} \right),
\nonumber \\
\vec{H}^{(i)}_{2} & =  & \frac{1}{i \mu_{i} k_{0}}  \vec \nabla \times \vec \nabla \times\left(
{U}^{(i)}  \vec{i}_{z} + {V}^{(i)}  \vec{r} \right);
\end{eqnarray}
(b) TM mode
\begin{eqnarray}
\vec{E}^{(i)}_{2} & = & - \frac{1}{i \varepsilon_{i} k_{0}}
\vec \nabla \times \vec \nabla \times \left({U}^{(i)}  \vec{i}_{z} + {V}^{(i)}  \vec{r} \right),
\nonumber \\
\vec{H}^{(i)}_{2} & = &
\vec \nabla \times \left( {U}^{(i)}  \vec{i}_{z} + {V}^{(i)}  \vec{r} \right),
\label{tm1}\end{eqnarray}
where the scalar potentials ${U}^{(i)}$ and ${V}^{(i)}$
satisfy the Helmholtz equations (\ref{wave}).

In the case of TE mode (see Eq.~(\ref{te0})), the boundary conditions for scalar
potentials have the form
\begin{equation} \begin{array}[t]{rl}
\left. \begin{array}{r}
\displaystyle  \eta U^{(j)} + \frac{d}{2} \xi  V^{(j)} =
\eta U^{(j+1)} + \frac{d}{2} \xi V^{(j+1)},
\\
\displaystyle  \frac{\partial{}}{\partial{\xi}} \left( \xi
 U^{(j)} + \tilde{f} \frac{d}{2} \eta V^{(j)} \right) = \frac{\partial{}}
{\partial{\xi}} \left( \xi U^{(j+1)} + \tilde{f} \frac{d}{2} \eta V^{(j+1)}
\right),
\\
\displaystyle  \varepsilon_{j} \left( \xi U^{(j)} + \tilde{f} \frac{d}{2} \eta V^{(j)}
\right) =
\varepsilon_{j+1} \left( \xi U^{(j+1)} + \tilde{f} \frac{d}{2} \eta V^{(j+1)}
\right),
\\
\displaystyle  \frac{1}{\mu_{j}} \frac{\partial{}}{\partial{\xi}} \left(
\eta U^{(j)} + \frac{d}{2} \xi V^{(j)}
\right) = \frac{1}{\mu_{j+1}} \left[ \frac{\partial{}} {\partial{\xi}}
\left( \eta U^{(j+1)} + \frac{d}{2} \xi V^{(j+1)} \right)
\right.
\\
\displaystyle + \left.
\left( 1 -
\frac{ c_{j+1}^{2}}{ c^{2}_{j}} \right) \frac{1 - \eta^{2}}{\xi^{2} - \tilde{f}}
\frac{\partial{}}{\partial{\eta}} \left( \xi U^{(j+1)} + \tilde{f} \frac{d}{2}
\eta V^{(j+1)} \right) \right],
\end{array} \right\}
_{\xi = \xi_{j}} \end{array}
\end{equation}
where $j=2, 3, \ldots , N$.
The boundary conditions for the potentials
$U^{(0)}, U^{(1)}$ and $V^{(0)}, V^{(1)}$ can be written
in a similar way to the axisymmetric part (see Eqs.~(\ref{11}), (\ref{12})).

As in the case of the axisymmetric part,
we can derive the integral equations for the scalar potentials
$U^{(j)}$ and $V^{(j)}$
\bea & &
\frac{d}{2}\, (\xi^2_j - \tilde{f}) \int_{0}^{2 \pi} \int_{0}^{ \pi} \left\{
U^{(j+1)} \frac{\partial G_j} {\partial \xi^\prime} -
\frac{\mu_j}{\mu_{j+1}} \frac{\partial U^{(j+1)}} {\partial
\xi^\prime} G_j + \left( \frac{\varepsilon_{j+1}}{\varepsilon_{j}} -
1 \right) \left[ \frac{\xi_j^2}{\xi_j^2 - \tilde{f} \eta^{\prime^2}}
\, U^{(j+1)}
\right. \right. \nonumber \\ \nonumber \\ & & \left.
\left. +
 \frac{ \tilde{f} \xi_j \eta^\prime}{\xi_j^2 - \tilde{f} \eta^{\prime^2}}
\, \frac{d}{2}\, V^{(j+1)} \right]
\frac{\partial G_j} {\partial \xi^\prime} +
\left( \frac{\mu_{j}}{\mu_{j+1}} -1 \right)
\left[ \frac{\xi_j^2}{\xi_j^2 - \tilde{f} \eta^{\prime^2}}
\frac{\partial{U^{(j+1)}}}{\partial{\xi^\prime}} +
 \frac{ \tilde{f} \xi_j \eta^\prime}{\xi_j^2 - \tilde{f} \eta^{\prime^2}} \, \frac{d}{2} \,
\frac{\partial{V^{(j+1)}}}{\partial{\xi^\prime}}
\right] G_j \right.
\nonumber \\ \nonumber \\ & & \left. +
\left( \frac{\varepsilon_{j+1}}{\varepsilon_{j}} - 1 \right)
\frac{\xi_j}{\xi_j^2 - \tilde{f} \eta^{\prime^2}}
\left[-U^{(j+1)} + \frac{2 \xi_j^2}{\xi_j^2 - \tilde{f} \eta^{\prime^2}} \, U^{(j+1)} +
 \frac{2 \tilde{f} \xi_j \eta^\prime}{\xi_j^2 - \tilde{f} \eta^{\prime^2}}
\, \frac{d}{2} \, V^{(j+1)} \right] G_j  \right.
\nonumber \\ \nonumber \\  & & \left. -
\left( \frac{\varepsilon_{j+1}}{\varepsilon_{j}} -
\frac{\mu_{j}}{\mu_{j+1}} \right)
 \frac{\tilde{f} \eta^\prime}{\xi_j^2 - \tilde{f} \eta^{\prime^2}}
 \left[ \frac{1 - \eta^{\prime^2}}{\xi_j^{2} - \tilde{f}}
\frac{\partial{}}{\partial{\eta^\prime}} \left( \xi_j \, U^{(j+1)} + \tilde{f}
\eta^\prime \, \frac{d}{2} \, V^{(j+1)} \right) + \frac{d}{2} \, V^{(j+1)}
\right] G_j \right \} {\rm d } \eta^\prime {\rm d } \varphi^\prime
\nonumber \\ \nonumber \\  & &
= \left\{ \ba{lcl} -U^{(j)}_{\rm A}
(\vec{r}), \ \ \ \ \ \   \vec{r} \in { D_j},
\\
U^{(j)}_{\rm B} (\vec{r}), \ \ \   \vec{r} \in R^3 \setminus \bar{ D_j},
\ea \right.
\label{39}
\eea
\bea
& &
 \frac{d}{2} \, (\xi^2_j - \tilde{f}) \int_{0}^{2 \pi} \int_{0}^{ \pi}
\left\{ \frac{\varepsilon_{j+1}}{\varepsilon_{j}} \,\frac{d}{2}\, V^{(j+1)}
\frac{\partial G_j}
{\partial \xi^\prime} -
\frac{d}{2}\, \frac{\partial V^{(j+1)}}
{\partial \xi^\prime} G_j -
\left( \frac{\varepsilon_{j+1}}{\varepsilon_{j}} - 1 \right)
\left[ \frac{\xi_j \eta^\prime}{\xi_j^2 - \tilde{f} \eta^{\prime^2}} \, U^{(j+1)}
\right. \right.
\nonumber \\ & & \left. \left.
\right. \right. \nonumber \\ & & \left. \left. +
 \frac{ \xi_j^2}{\xi_j^2 - \tilde{f} \eta^{\prime^2}} \,
\frac{d}{2}\, V^{(j+1)} \right]
\frac{\partial G_j} {\partial \xi^\prime} -
\left( \frac{\mu_{j}}{\mu_{j+1}} -1 \right)
\left[ \frac{\xi_j \eta^\prime}{\xi_j^2 - \tilde{f} \eta^{\prime^2}}
\frac{\partial{U^{(j+1)}}}{\partial{\xi^\prime}} +
 \frac{ \xi_j^2}{\xi_j^2 - \tilde{f} \eta^{\prime^2}} \,\frac{d}{2}\,
\frac{\partial{V^{(j+1)}}}{\partial{\xi^\prime}}
\right] G_j \right.
\nonumber \\ \nonumber \\ & & \left. -
\left( \frac{\varepsilon_{j+1}}{\varepsilon_{j}} - 1 \right)
\frac{\xi_j}{\xi_j^2 - \tilde{f} \eta^{\prime^2}}
\left[- \frac{d}{2}\, V^{(j+1)} +
\frac{2 \xi_j \eta^{\prime}}{\xi_j^2 - \tilde{f} \eta^{\prime^2}} \, U^{(j+1)} +
 \frac{2 \xi_j^2}{\xi_j^2 - \tilde{f} \eta^{\prime^2}}\,
\frac{d}{2} \, V^{(j+1)} \right] G_j \right.
\nonumber \\ \nonumber \\ & & \left. +
\left( \frac{\varepsilon_{j+1}}{\varepsilon_{j}} -
\frac{\mu_{j}}{\mu_{j+1}} \right)
 \frac{ \xi}{\xi_j^2 - \tilde{f} \eta^{\prime^2}}
 \left[ \frac{1 - \eta^{\prime^2}}{\xi_j^{2} - \tilde{f}}
\frac{\partial{}}{\partial{\eta^\prime}} \left( \xi_j \, U^{(j+1)} +
\tilde{f} \eta^\prime \,
\frac{d}{2} \, V^{(j+1)} \right) + \frac{d}{2} \, V^{(j+1)} \right] G_j
\right \} {\rm d } \eta^\prime {\rm d } \varphi^\prime
\nonumber \\ \nonumber \\  & & =
\left\{
\ba{lcl}
-V^{(j)}_{\rm A} (\vec{r}), \ \ \ \ \ \   \vec{r} \in { D_j},
\\
V^{(j)}_{\rm B} (\vec{r}), \ \ \   \vec{r} \in R^3 \setminus \bar{ D_j},
\ea \right.
\label{40}
\eea
where $\xi^\prime = \xi_j$.

The scalar potentials are expanded in terms of the
spheroidal functions \cite{KPS76}
\be \ba{l}
U^{(0)} \\
V^{(0)}
\ea
= \sum_{m=1}^\infty\ \sum_{l=m}^\infty\
\ba{l}
a^{(0)}_{ml} \\
b^{(0)}_{ml}
\ea
R_{ml}^{(1)} (c_1, \xi)\ S_{ml} (c_1, \eta)\ \cos m\varphi,
\label{uv10}\ee
\be
\ba{l}
U^{(1)} \\
V^{(1)}
\ea
= \sum_{m=1}^\infty\ \sum_{l=m}^\infty\
\ba{l}
a^{(1)}_{ml} \\
b^{(1)}_{ml}
\ea
R_{ml}^{(3)} (c_1, \xi)\ S_{ml} (c_1, \eta)\ \cos m\varphi,
\label{uv20}\ee
\be
\ba{l}
U^{(j)}_{\rm A} \\
V^{(j)}_{\rm A}
\ea
= \sum_{m=1}^\infty\ \sum_{l=m}^\infty\
\ba{l}
a^{(j)}_{ml} \\
b^{(j)}_{ml}
\ea
R_{ml}^{(1)} (c_j, \xi)\ S_{ml} (c_j, \eta)\ \cos m\varphi,
\label{uv1}\ee
\be
\ba{l}
U^{(j)}_{\rm B} \\
V^{(j)}_{\rm B}
\ea
= \sum_{m=1}^\infty\ \sum_{l=m}^\infty\
\ba{l}
c^{(j)}_{ml} \\
d^{(j)}_{ml}
\ea
R_{ml}^{(3)} (c_j, \xi)\ S_{ml} (c_j, \eta)\ \cos m\varphi,
\label{uv2}\ee
where $j=2, 3, \ldots , N+1$.
For the TE mode, the coefficients that describe the incident radiation
are equal to (see \cite{vf93, fvs96})
\be
{a}^{(0)}_{ml} = \frac{4 i^{l-1}}{k_{1}} N^{-2}_{ml}(c_{1})
\frac{S_{ml}(c_{1},\cos{\alpha})}{\sin{\alpha}}\,,\,\,\,
{b}^{(0)}_{ml} = 0.
\ee
For TM mode, the coefficients ${a}^{(0)}_{ml}$ have the opposite
sign and the multiplicand $\sqrt{ { \varepsilon_{1} }/{ \mu_{1} }}$
(see  Eqs.~(\ref{tm0}), (\ref{tm1})).

Substituting Eqs.~(\ref{uv10})--(\ref{uv2}), and (\ref{28}) into
integral equations (\ref{39}) and (\ref{40}), we obtain infinite systems
relative to the unknown expansion coefficients. The systems
can be written in the matrix form ($j=1, 2, \ldots , N$)
\be
 \left( \begin{array}{c}
\vec{Z}^{(j)}_{\rm A} \\
\vec{Z}^{(j)}_{\rm B}
          \end{array} \right) =
{-{\cal A}^{(j)}_{31}  -{\cal A}^{(j)}_{33}
\choose
  {\cal A}^{(j)}_{11} \ \ \ \ \ \  {\cal A}^{(j)}_{13}}
 \left( \begin{array}{c}
\vec{Z}^{(j+1)}_{\rm A} \\
\vec{Z}^{(j+1)}_{\rm B}
          \end{array} \right),
\label{sys}\ee
where the vectors and matrices have the block structure
\be
 \left( \begin{array}{c}
\vec{Z}^{(j)}_{\rm A} \\
\vec{Z}^{(j)}_{\rm B}
          \end{array} \right) =
 \left( \begin{array}{c}
         \left\{ k_{1} {a}^{(j)}_{ml} R^{(3)}_{ml}(c_{j}, \xi_{j}) N_{ml}(c_{j}) \right\}^{\infty}_{m\phantom{_k}} \\
         \left\{ c_{1} {b}^{(j)}_{ml} R^{(3)}_{ml}(c_{j}, \xi_{j}) N_{ml}(c_{j}) \right\}^{\infty}_{m\phantom{_k}} \\
         \left\{ k_{1} {c}^{(j)}_{ml} R^{(3)}_{ml}(c_{j}, \xi_{j}) N_{ml}(c_{j}) \right\}^{\infty}_{m\phantom{_k}} \\
         \left\{ c_{1} {d}^{(j)}_{ml} R^{(3)}_{ml}(c_{j}, \xi_{j}) N_{ml}(c_{j}) \right\}^{\infty}_{m}
          \end{array} \right),
\ee
\be
{\cal A}^{(j)}_{ik}
=
{ \mathfrak{A}^{(j)}_{ik,{\rm A}} \ \ \  \mathfrak{B}^{(j)}_{ik,{\rm A}}
\choose
  \mathfrak{A}^{(j)}_{ik,{\rm B}} \ \ \  \mathfrak{B}^{(j)}_{ik,{\rm B}} },
\label{aaa}\ee
\bea
\mathfrak{A}^{(j)}_{31,{\rm A}} & = &
{\cal W}_j \ \left\{ {\cal R}^{[3]}(c_j, \xi_j)\
\Delta^{(m)}(c_j, c_{j+1})
- \frac{\mu_j}{\mu_{j+1}}
\Delta^{(m)}(c_j, c_{j+1})\ {\cal R}^{[1]}(c_{j+1}, \xi_j)
\right.  \nonumber \\ & &
+ \left( \frac{\varepsilon_{j+1}}{\varepsilon_{j}} - 1 \right) \xi_j
\left[ \xi_j {\cal R}^{[3]}(c_j, \xi_j) Q^{(m)}(c_j, c_{j+1}, \xi_j)
 \right. \nonumber \\ & & \left. \left.
- Q^{(m)}(c_j, c_{j+1}, \xi_j)
\left( I - 2 \xi_j^2  Q^{(m)}(c_{j+1}, c_{j+1}, \xi_j)  \right) \right]
 \right. \nonumber \\ & & \left.
+ \left( \frac{\mu_j}{\mu_{j+1}} - 1 \right)\ \xi_j^2
Q^{(m)}(c_j, c_{j+1}, \xi_j){\cal R}^{[1]}(c_{j+1}, \xi_j)
\right. \nonumber \\ & & \left. -
\left( \frac{\varepsilon_{j+1}}{\varepsilon_{j}} -
\frac{\mu_j}{\mu_{j+1}} \right)\ \frac{\tilde{f} \xi_j}{\xi^2_j - \tilde{f}}\ Q^{(m)}(c_j, c_{j+1}, \xi_j)
E^{(m)}(c_{j+1}, c_{j+1})
\right\}
 \nonumber \\ & &
\times {\cal P}^{[1]}(c_{j+1}, \xi_j, \xi_{j+1}),
\label{a31}\eea
\bea
\mathfrak{B}^{(j)}_{31,{\rm A}} & = &
{\cal W}_j \ \left\{
\left( \frac{\varepsilon_{j+1}}{\varepsilon_{j}} - 1 \right)  \tilde{f} \xi_j
\left[ {\cal R}^{[3]}(c_j, \xi_j) Q^{(m)}(c_j, c_{j+1}, \xi_j)
\right. \right. \nonumber \\ & & \left. \left.
+ 2 \xi_j Q^{(m)}(c_j, c_{j+1}, \xi_j)  Q^{(m)}(c_{j+1}, c_{j+1}, \xi_j)
\right] \Gamma^{(m)}(c_{j+1}, c_{j+1})
\right. \nonumber \\ & & \left.
+ \left( \frac{\mu_j}{\mu_{j+1}} - 1 \right)\ \tilde{f} \xi_j
Q^{(m)}(c_j, c_{j+1}, \xi_j) \Gamma^{(m)}(c_{j+1}, c_{j+1}) {\cal R}^{[1]}(c_{j+1}, \xi_j) -
\left( \frac{\varepsilon_{j+1}}{\varepsilon_{j}} -
\frac{\mu_j}{\mu_{j+1}} \right)
\right. \nonumber \\ & & \left.
\times \frac{\tilde{f}}{\xi^2_j - \tilde{f}}\
\left[ ( \xi^2_j Q^{(m)}(c_j, c_{j+1}, \xi_j) - \Delta^{(m)}(c_j, c_{j+1})) K^{(m)}(c_{j+1}, c_{j+1}) +
 \Gamma^{(m)}(c_{j}, c_{j+1})  \right]
\right\} \
 \nonumber \\ & &
\times {\cal P}^{[1]}(c_{j+1}, \xi_j, \xi_{j+1}),
\label{b31}\eea
\bea
\mathfrak{A}^{(j)}_{31,{\rm B}} & = &
{\cal W}_j \ \left\{
- \left( \frac{\varepsilon_{j+1}}{\varepsilon_{j}} - 1 \right)   \xi_j
\left[ {\cal R}^{[3]}(c_j, \xi_j) Q^{(m)}(c_j, c_{j+1}, \xi_j)
\right. \right. \nonumber \\ & & \left. \left.
+ 2 \xi_j Q^{(m)}(c_j, c_{j+1}, \xi_j)
 Q^{(m)}(c_{j+1}, c_{j+1}, \xi_j) \right] \Gamma^{(m)}(c_{j+1}, c_{j+1})
\right. \nonumber \\ & & \left.
- \left( \frac{\mu_j}{\mu_{j+1}} - 1 \right)\  \xi_j
Q^{(m)}(c_j, c_{j+1}, \xi_j) \Gamma^{(m)}(c_{j+1}, c_{j+1}) {\cal R}^{[1]}(c_{j+1}, \xi_j) +
\left( \frac{\varepsilon_{j+1}}{\varepsilon_{j}} -
\frac{\mu_j}{\mu_{j+1}} \right)\
 \right. \nonumber \\ & & \left.
\times \frac{ \xi^2_j}{\xi^2_j - \tilde{f}}\
 Q^{(m)}(c_j, c_{j+1}, \xi_j) K^{(m)}(c_{j+1}, c_{j+1})
\right\} \ {\cal P}^{[1]}(c_{j+1}, \xi_j, \xi_{j+1}),
\label{a32}\eea
\bea
\mathfrak{B}^{(j)}_{31,{\rm B}} & = &
{\cal W}_j \ \left\{ \frac{\varepsilon_{j+1}}{\varepsilon_{j}}
{\cal R}^{[3]}(c_j, \xi_j)\ \Delta^{(m)}(c_j, c_{j+1}) -
\Delta^{(m)}(c_j, c_{j+1})\ {\cal R}^{[1]}(c_{j+1}, \xi_j)
 \right. \nonumber \\ & & \left.
- \left( \frac{\varepsilon_{j+1}}{\varepsilon_{j}} - 1 \right) \xi_j
\left[ \xi_j {\cal R}^{[3]}(c_j, \xi_j) Q^{(m)}(c_j, c_{j+1}, \xi_j)
\right. \right. \nonumber \\ & & \left. \left.
- Q^{(m)}(c_j, c_{j+1}, \xi_j)
(I - 2 \xi_j^2  Q^{(m)}(c_{j+1}, c_{j+1}, \xi_j) ) \right]
 \right. \nonumber \\ & & \left.
- \left( \frac{\mu_j}{\mu_{j+1}} - 1 \right)\ \xi_j^2
Q^{(m)}(c_j, c_{j+1}, \xi_j) {\cal R}^{[1]}(c_{j+1}, \xi_j)
 \right. \nonumber \\ & & \left.
 + \left( \frac{\varepsilon_{j+1}}{\varepsilon_{j}}
- \frac{\mu_j}{\mu_{j+1}} \right)\ \frac{\xi_j}{\xi^2_j - \tilde{f}}\
\left[\tilde{f} Q^{(m)}(c_j, c_{j+1}, \xi_j) E^{(m)}(c_{j+1}, c_{j+1})
+ \Delta^{(m)}(c_j, c_{j+1}) \right]
\right\} \
 \nonumber \\ & &
\times {\cal P}^{[1]}(c_{j+1}, \xi_j, \xi_{j+1}),
\label{b32}\eea
\be
Q^{(m)}(c_{j+1}, c_{j+1}, \xi_j) = {\left\{ \xi_j^{2} I - \tilde{f} \left[\Gamma^{(m)}(c_{j+1}, c_{j+1})\right]^2 \right\}}^{-1},
\ee
\be
Q^{(m)}(c_j, c_{j+1}, \xi_j) =  \Delta^{(m)}(c_j, c_{j+1})\, Q^{(m)}(c_{j+1}, c_{j+1}, \xi_j)
\ee
and $I = \left\{ \delta_{nl} \right\}^{\infty}_{m}$ is the unit matrix.
Here, $m$ is the azimuthal index that runs from unity
to infinity. The subscripts and superscripts of the matrices
have the same meaning as in the previous section.
The elements of the remaining matrices in Eq.~(\ref{aaa})
can be obtained from Eqs.~(\ref{a31})--(\ref{b32}) as explained
above (see Eqs.~(\ref{a0-31})--(\ref{a0-13})).
The matrix elements
$ \Gamma^{(m)}(c_j, c_j) = \left\{ \gamma^{(m)}_{nl} (c_{j}, c_{j})
\right\}^{\infty}_{m}$,
$K^{(m)}(c_j, c_j) = \left\{ \kappa^{(m)}_{nl} (c_{j}, c_{j})
\right\}^{\infty}_{m}$ and
$E^{(m)}(c_j, c_j) = \left\{ \varepsilon^{(m)}_{nl} (c_{j}, c_{j}) \right\}^{\infty}_{m}$
are the integrals of products of the angular
spheroidal functions and their derivatives
(see \cite{vf93, fvs96}).

The system (\ref{sys}) can be easily solved for
the expansion coefficients of the scattered radiation
potential (cf. Eq.~(\ref{26}))
\be
\vec{Z}^{(1)}_{\rm B} = {\cal A}_2  {\cal A}_1^{(-1)} \ \vec{Z}^{(1)}_{\rm A},
\ee
where
\be
\vec{Z}^{(1)}_{\rm B} = \left( \begin{array}{c}
         \left\{ k_{1} {a}^{(1)}_{ml} R^{(3)}_{ml}(c_{1}, \xi_{1}) N_{ml}(c_{1}) \right\}^{\infty}_{m\phantom{_k}} \\
         \left\{ c_{1} {b}^{(1)}_{ml} R^{(3)}_{ml}(c_{1}, \xi_{1}) N_{ml}(c_{1})      \right\}^{\infty}_{m\phantom{_k}}
          \end{array} \right),
\ee
\be
\vec{Z}^{(1)}_{\rm A} = \left( \begin{array}{c}
                \left\{ k_{1} {\it a}^{(0)}_{ml} R^{(1)}_{ml}(c_{1}, \xi_{1})\, N_{ml}(c_{1})  \right\}^{\infty}_{m} \\
                 0
                  \end{array} \right),
\ee
and the matrices ${\cal A}_1$ and ${\cal A}_2$
satisfy Eq.(\ref{matr}) but have the block structure (see Eq.~(\ref{aaa})).

For the TM mode, we can obtain infinite
systems for the unknown expansion coefficients of the
scalar potentials by replacing
$\mu_j \rightarrow \varepsilon_j, \, \, \varepsilon_j \rightarrow \mu_j$
in the above relations.
Note that in the case of $\mu_j = 1 $,
their form is much simpler than in the corresponding case of
the TE mode (see also \cite{vf93, fvs96}).

\section{Characteristics of Scattered Radiation}

Using the expansion coefficients of the scattered field for
the TE and TM polarizations,
we can calculate elements of the scattering matrix (see the corresponding
expressions in \cite{fvs96}) and
the integral characteristics of the scattered radiation
(e.g., the cross sections of a particle for extinction $C_{\rm ext}$,
scattering  $C_{\rm sca}$, absorption $C_{\rm abs}$, and
radiation pressure  $C_{\rm pr}$).
These cross sections are products of the corresponding efficiency
factors $Q$ and  the viewing geometric cross section of a spheroid (the area
of the particle shadow)
$$
C = G Q,
$$
where
\begin{equation}
G(\alpha) = \pi b_1 \left(a_1^2\sin^2\alpha
            + b_1^2\cos^2\alpha\right)^{1/2}
             \,\,\,\,\,\,\, \mbox{for prolate  spheroids,}
\end{equation}
\begin{equation}
G(\alpha) = \pi a_1\left(a_1^2\cos^2\alpha
            + b_1^2\sin^2\alpha\right)^{1/2}
             \,\,\,\,\,\, \mbox{for oblate spheroids}
\end{equation}
and $a_1$ and $b_1$ are the major and minor semiaxes of a multilayered spheroid.
The efficiency factors for extinction can be found as
\begin{eqnarray}
& Q_{\rm ext} &\,\, =  \frac{4}{
c^{2}_{1}
{\left[ ( \xi^{2}_{1} - \tilde{f} ) ( \xi^{2}_{1} - \tilde{f} {\cos}^{2}{\alpha} ) \right]}^{1/2}
}
\, {\rm Re} \left[ -
\sum^{\infty}_{l=1} i^{-l} {\it a}^{(1)}_{l} S_{1l}(c_{1}, \cos{\alpha})
 \right. \nonumber \\
&\!\!\! + & \sum^{\infty}_{m=1} \sum^{\infty}_{l=m} i^{-(l-1)}
\Bigl(
k_{1} {\it a}^{(1)}_{ml} S_{ml}(c_{1}, \cos{\alpha}) +
\left. \left. i {\it b}^{(1)}_{ml} S^{'}_{ml}(c_{1}, \cos{\alpha}) \right)
\sin{\alpha} \Biggr]  \right. ,
\end{eqnarray}
where, as above,
$\tilde{f} = 1$ for prolate spheroids and $\tilde{f} = - 1$ for oblate ones.
Expressions for other factors are given in \cite{vf93, fvs96}.

To compare the optical properties of particles of various
shape, the cross sections can be normalized by the geometric cross section
of the equivolume sphere
\be
\frac{C}{\pi r^{2}_{V}} = \frac{[(a_1/b_1)^2\sin^2\alpha
                             +\cos^2\alpha]^{1/2}}{(a_1/b_1)^{2/3}} \, Q
\,\,\,\,\,\,\,\,\,\,\, \mbox{for prolate  spheroids,}
\label{crv_pro}\ee
\be
\frac{C}{\pi r^{2}_{V}} = \frac{[(a_1/b_1)^2\cos^2\alpha
                             +\sin^2\alpha]^{1/2}}{(a_1/b_1)^{1/3}} \, Q
             \,\,\,\,\,\,\,\,\,\,\, \mbox{for oblate  spheroids.}
\label{crv_obl}
\ee
Here, $r_{V}$ is the radius of the sphere with the volume equal to
that of a given spheroidal particle.
This radius can be defined as
\begin{equation}
r_V^{3} = a_1 \, b^{2}_1
             \,\,\,\,\,\,\,\,\,\,\, \mbox{for prolate  spheroids,}
\end{equation}
\begin{equation}
r_V^{3} = a^{2}_1 \, b_{1}
             \,\,\,\,\,\,\,\,\,\,\, \mbox{for oblate  spheroids.}
\end{equation}

The optical properties of a multilayered confocal spheroid
can be found if we put the type of spheroid (prolate or
oblate), the number of layers $N$, complex refractive indices of all layers
$m_j=n_j + k_j i$, the outer aspect ratio
$a_1/b_1$ ($a_1$  and $b_1$ are the major and minor semiaxes),
the total particle size parameter,
and the relative ratios of volumes of the layers $f_j= V_{j} /V_{\rm total}$.
The size parameter may be specified as
$$
x_V = 2 \pi r_V/ \lambda,
$$
where $r_V$ is the radius of a sphere whose volume is equal to that of
the spheroid, $\lambda$ the wavelength of incident radiation.

The radial coordinates $\xi_{j}$ that define
the boundaries of a layered particle, are connected with the
corresponding semiaxes as
\begin{equation}
\xi_{j} = \left(\frac{a_{j}}{b_{j}}\right)^{(1+\tilde{f})/2} \left[\left(
            \frac{a_{j}}{b_{j}}\right)^{2} - 1 \right]^{-1/2}.
\end{equation}
The efficiency factors can also be considered as a function of the size
parameter $2 \pi a_{1} / \lambda$ given by
\begin{equation}
\frac{2 \pi a_{1}}{\lambda} = \left(\frac{a_{1}}{b_{1}}\right)^{(1-\tilde{f})/2} \,c_{1} \xi_{1}
= x_V\, \left(\frac{a_{1}}{b_{1}}\right)^{(3+\tilde{f})/6}.
\label{s2}\end{equation}

The ratio of layer volumes inside the surface $j$
($j=2,3, \ldots , N$)
to the total volume of a multilayered particle
is determined by using the parameters $\xi_{1}$ and $\xi_{j}$
\begin{equation}
\sum_{l=j}^N\,{f_{l}} =
\sum_{l=j}^N\,\frac{V_{l}}{V_{\rm total}} =
\frac{\xi_{j}\,(\xi^{2}_{j} - \tilde{f})}
{\xi_{1}\,(\xi^{2}_{1} - \tilde{f})}.
\end{equation}
The aspect ratios of internal layers can be calculated from
the volume ratios
by using the iterative procedure for prolate spheroids
\begin{equation}
(\xi_{j})^{(n)} = \sqrt[3]{ (\xi_{j})^{(n-1)} +
\sum_{l=j}^N\,{f_{l}}
\,[\xi_{1}\,(\xi_{1}^{2} - 1)] } ,
\end{equation}
where $n=1,\,2\, \ldots$ and the initial value $(\xi_{j})^{(0)} = \xi_{j-1}$.
For oblate spheroids, the
parameter $\xi_{j}$ can be found by Newton's method
\begin{equation}
(\xi_{j})^{(n)} = \frac{2 \left[ (\xi_{j})^{(n-1)} \right ]^3 +
\sum_{l=j}^N\,{f_{l}}
                    \,[\xi_{1}\,(\xi_{1}^{2} + 1)] }
                    {3 \left[ (\xi_{j})^{(n-1)} \right ]^2 + 1},
\end{equation}
where $n=1,\,2\, \ldots$ and the initial value $(\xi_{j})^{(0)} = 0$.

\section{Numerical Results and Discussion}

\subsection{Computational Tests}

The created computer code is
based on our codes developed earlier for homogeneous \cite{vf93} and coated
spheroids \cite{fvs96}.
In calculations of the radial spheroidal functions,
we use their expansions in terms of the  Legendre or Bessel functions,
the solution to the corresponding differential equation, or J\'affe expansion
for prolate functions according to the recommendations
given in \cite{vf02, vf04}.

The numerical code has been examined by using various tests that include
internal control (see Table~\ref{t1}),
a comparison with the known results for homogeneous and core-mantle
spheroids and multilayered spheres \cite{vm99} (Fig.~\ref{ss})
as well as a comparison with the calculations for multilayered
spheroids based on
the EBCM with a spherical basis \cite{fip03},
the SVM with a spherical basis \cite{vfi09},
and the quasistatic approximation \cite{Po_02}.
We also have considered absorbing and dielectric particles with a different
number of layers and various aspect ratios.

For non-absorbing particles the efficiency factors for extinction
and scattering are known to be equal for the same  azimuthal index $m$:
$Q^{(m)}_{\rm ext} = Q^{(m)}_{\rm sca}$ ($Q = \sum_{m} Q^{(m)}$).
Then by increasing the number of terms $N_{\max}$ in sums
for $Q^{(m)}_{\rm ext}$ and $Q^{(m)}_{\rm sca}$,
one should  obtain
a decreasing difference between these two factors,
i.e., $\left| Q^{(m)}_{\rm ext} - Q^{(m)}_{\rm sca}\right| \rightarrow 0$,
if $N_{\max} \rightarrow \infty$.
Table~\ref{t1} shows the behavior of the efficiency
factors in the case of radiation propagating along the rotation axis of
a spheroid when the sums over $m$ contain only one term, $m= 1$.
A comparison with the results presented in \cite{fvs96}
demonstrates that the convergence for multilayered spheroids resembles
that for coated spheroids, i.e, it is not a function of the number of layers $N$.
The convergence is determined by the particle size $2\pi a_1/\lambda$
and is independent of its shape.
The latter feature makes our solution with a spheroidal basis
essentially different from the SVM or EBCM approach with spherical basis
when convergence quickly degrades for spheroids
with aspect ratios $a_1/b_1 \ga 1.5 - 2$ \cite{fip03, vfi09}.
Note that our code allows one to calculate the optical properties
of spheroids with the size parameters up to
$x_V \approx 15 - 20$ including very elongated or flattened particles.
The time of calculations grows with the increase
of layers number as $t \approx N^{1.2}$, which is much faster
as compared to other methods  ($t \approx N^{2.5-3}$ see
discussion in \cite{vfi09}).

If the particles are nearly spherical, the optical properties
of multilayered spheroids and  spheres should be almost the same.
We have considered various absorbing and dielectric particles and
some results are shown in Fig.~\ref{ss} where the relative differences in
percents
\begin{equation}
\epsilon = \frac{Q_{\rm ext}({\rm sphere}) - C_{\rm ext}({\rm spheroid})
           /\pi r_{V}^2}
{Q_{\rm ext}({\rm sphere})} \, 100 \%
\label{eps}
\end{equation}
are given. The values of $\epsilon$ are plotted as a function of
the size parameter $x_V$ for particles with the aspect ratio
$a_1/b_1 = 1.0001$ and the volume ratios
$V_j/V_{\rm total} = 0.33$ ($j=1, 2, 3$).
The aspect ratios of internal layers are equal to
$a_2/b_2 \approx 1.00016$ and
$a_3/b_3 \approx 1.00021$. The wavelike behavior is typical only
for dielectric particles. For highly absorbing particles, the values
of $\epsilon$ demonstrate a smooth, monotonous growth with increasing $x_V$.


\subsection{Particles with a Different Number of Layers}

The model of multilayered spheroids gives wide opportunities
to investigate both the shape and structure effects
on the optics  of composite particles simultaneously.
As one of the  first applications of the developed model,
is our analysis of the idea to represent
some composite interstellar grains by multilayered
particles  as suggested in  \cite{vm99}.
We consider multilayered spheroids with
different material layers cyclically changing inside a particle.
The particles are assumed to be composed of
amorphous carbon or silicate with varied volume fraction of vacuum.
The chosen optical constants for carbon ($m=1.98+0.23i$) and silicate
($m=1.68+0.03i$) correspond to the wavelength $\lambda=0.55\,\mu$m.

The extinction efficiencies of equivolume layered prolate spheroids
with  $a_1/b_1 = 3$ are compared in Fig.~\ref{f3}.
In this case, the aspect
ratio of the innermost layer is equal to
$a_3/b_3 = 4.91$, $a_9/b_9 = 8.33$ and $a_{18}/b_{18} = 11.72$ for
3-layered, 9-layered and 18-layered particles, respectively.

The size parameters  of compact and porous particles are related as
\be
x_{\rm porous} = \frac{x_{\rm compact}}{(1-{\cal P})^{1/3}}
= \frac{x_{\rm compact}}{(V_{\rm solid} /V_{\rm total})^{1/3}}, \label{xpor}
\ee
where the particle porosity ${\cal P}$ ($0 \leq {\cal P} < 1$)
is introduced as
$
{\cal P} = V_{\rm vac} /V_{\rm total}
= 1 - V_{\rm solid} /V_{\rm total}
$
and $V_{\rm vac}$ and $V_{\rm solid}$ are the volume fractions
of vacuum and solid material, respectively.

As in the case of layered spheres (see \cite{vm99,vih04}),
the scattering characteristics of
layered spheroids slightly depend on the order of materials and
become close
to some ``average'' ones, when particles consist of many layers
($\ga 15 - 20$).

The convergence of the extinction factors seems to be  better for oblique
incidence and oblate particles and larger aspect ratios.
Such a behavior is typical for
other efficiencies (scattering, absorption), albedo and the asymmetry
parameter.

\subsection{Particles of Different Porosity}

The difference in the optical properties of compact and porous particles
is clearly seen in Fig.~\ref{f4} which
shows the size dependence of normalized cross sections as given
by Eqs.~(\ref{crv_pro}), (\ref{crv_obl}).
The compact and porous particles have the same mass
for the same size parameter. This means that variations of the extinction
are related to the changes in the particle shape, orientation, porosity, and
the particle  type (prolate or oblate).
As follows from Fig.~\ref{f4}, the position of the first
maximum shifts to larger size parameters with a growth of porosity.
For very large particles, the normalized cross sections of compact
and porous particles cease to fluctuate and become rather similar.

The role of porosity in dust optics can be properly analyzed by using
the normalized cross sections
\bea
C^{\rm (n)} = \frac{C({\rm porous \, grain})}
{C({\rm  compact \, grain \, of \, same \, mass \, and \, shape})}
\nonumber \\ \,\,\,\,\,\,\,
=  (1-{\cal P})^{-2/3}\,  \frac{Q({\rm porous \, grain})}
{Q({\rm  compact \, grain \, of \, same \, mass \, and \, shape})}. \label{cn}
\eea
The quantity $C^{\rm (n)}$ shows how porosity can increase or decrease
the cross section.
Such an investigation was performed in \cite{vih04,ks94} for spherical
particles.

Figures~\ref{f6}, \ref{f7} show the extinction
cross sections $C^{\rm (n)}$ computed
for prolate and oblate spheroids with porosity ${\cal P}=0.33$ and 0.5.
It is seen that the behavior of curves
$C^{\rm (n)} (x_{\rm compact})$ is rather complicated.
The  porosity increases the extinction for spheroids
of almost all sizes and shapes that are
seen pole-on ($\alpha=0\degr$) and
decreases the extinction for spheroids  that are seen edge-on
($\alpha=90\degr$), if $x_{V,{\rm compact}}\la 3 - 5$. Note that in
the last case the curves are plotted for the sum of the TM and TE modes
for non-polarized incident radiation.
For very large particles,
the normalized cross sections tends to approach to asymptotic values
$C^{\rm (n)}  \rightarrow (1-{\cal P})^{-2/3}$
(see Eq.~(\ref{cn})) which are equal to 1.31 and 1.59 if
${\cal P} =0.33$ and 0.5, respectively.

\section{Conclusions}\label{concl}

The main results of the paper are as follows:

1. We have solved the light scattering problem for a
confocal multilayered spheroid by using the extended boundary condition
method (EBCM) with a corresponding spheroidal basis.
Our recursive solution is based on a special choice of scalar potentials
which for any next layer can be found by using the potentials of the
previous layer, the procedure starting from the particle core.

2. The numerical code has been thoroughly examined by using various tests.
They have demonstrated that the convergence of the efficiency factors
for multilayered spheroids is not a function of layers number $N$
and is independent of particle shape.
These features make our solution with a spheroidal basis
essentially different from the SVM or EBCM approach with spherical basis.
Our code allows one to calculate the optical properties
of spheroids with the size parameters up to
$x_V = 2 \pi r_V/ \lambda \approx 15 - 20$
($r_V$ is the radius of equivolume sphere)
including very elongated or flattened particles.

3. Illustrative calculations show the convergence of the extinction factors
to average values with a growing number of layers.
In the case of the large number of layers,
the optical properties of layered particles
slightly depend on the order of the materials and are determined by
the volume fraction of the materials.
Variations of extinction with a growth of particle porosity
demonstrate increase of the extinction for spheroids
of almost all sizes and shapes  that are seen pole-on ($\alpha=0\degr$) and
decrease of the extinction for spheroids  that are seen edge-on
($\alpha=90\degr$), if $x_{V,{\rm compact}}\la 3 - 5$.

\section*{Acknowledgments}
We thank Marina Prokopjeva and Alexander Vinokurov for test calculations
and Vladimir Il'in 
for helpful comments.
The work was partly supported by
grants  RFBR 10-02-00593 and 11-02-92695.



\newpage
\clearpage
\setcounter{table}{0}

\section*{List of Table Captions}

Table 1.
Efficiency factors for Extinction $Q_{\rm ext}$
and Scattering  $Q_{\rm sca}$ for Prolate and Oblate Multilayered
Spheroids at $\alpha = 0^{\circ}$.

\clearpage
\newpage
\begin{sidewaystable}[ht]
\bc
\caption{\mbox{Efficiency factors for Extinction $Q_{\rm ext}$
and Scattering  $Q_{\rm sca}$ for Prolate and Oblate Multilayered
Spheroids at $\alpha = 0^{\circ}$ $^a$} }\label{t1}
\begin{tabular}{cllcllcllcll}
\hline
\multicolumn{1}{c}{} &
\multicolumn{5}{c}{Prolate spheroid} &&
\multicolumn{5}{c}{Oblate spheroid} \\
\multicolumn{1}{c}{} &
\multicolumn{2}{c}{$a_1/b_1$ = 2} &&
\multicolumn{2}{c}{$a_1/b_1$ = 10} &&
\multicolumn{2}{c}{$a_1/b_1$ = 2} &&
\multicolumn{2}{c}{$a_1/b_1$ = 10} \\
   \cline{2-3}
   \cline{5-6}
   \cline{8-9}
   \cline{11-12}
\multicolumn{1}{c}{~$N_{\max}$} &
\multicolumn{1}{c}{$Q_{\rm ext}$} &
\multicolumn{1}{c}{$Q_{\rm sca}$} &&
\multicolumn{1}{c}{$Q_{\rm ext}$} &
\multicolumn{1}{c}{$Q_{\rm sca}$} &&
\multicolumn{1}{c}{$Q_{\rm ext}$} &
\multicolumn{1}{c}{$Q_{\rm sca}$} &&
\multicolumn{1}{c}{$Q_{\rm ext}$} &
\multicolumn{1}{c}{$Q_{\rm sca}$} \\
   \hline
~6    & 7.3        & 7.9        && 0.35         & 0.41         && 2.30         & 2.26         && 0.26         & 0.23          \\
~8    & 7.40       & 7.36       && 0.330        & 0.332        && 2.409        & 2.413        && 0.252        & 0.251         \\
10    & 7.386      & 7.387      && 0.3267       & 0.3264       && 2.4108       & 2.4105       && 0.2544       & 0.2542        \\
12    & 7.38691    & 7.38688    && 0.32680      & 0.32679      && 2.410809     & 2.410812     && 0.25426      & 0.25427       \\
14    & 7.3869017  & 7.3869022  && 0.3268027    & 0.3268029    && 2.4108093    & 2.4108092    && 0.254276     & 0.254275      \\
16    & 7.38690174 & 7.38690174 && 0.326802788  & 0.326802792  && 2.410809320  & 2.410809320  && 0.25427511   & 0.25427512    \\
18    & 7.38690174 & 7.38690174 && 0.3268027850 & 0.3268027850 && 2.4108093212 & 2.4108093212 && 0.2542751277 & 0.2542751273  \\
   \hline \\
\end{tabular}
\ec
\footnotesize $^a$ the number of layers $N=18$,
${m}_{j} = 1.3+0.0i$, $j=1,4,7,10,13,16$;
${m}_{j} = 1.5+0.0i$, $j=2,5,8,11,14,17$;
${m}_{j} = 1.7+0.0i$, $j=3,6,9,12,15,18$;
$2\pi a_1/\lambda = 5$,  and
$V_j/V_{\rm total} = 1/18$.
\global\def\rotfloatpage{R}
\end{sidewaystable}

\newpage
\clearpage
\setcounter{figure}{0}

\section*{List of Figure Captions}


Fig. 1.
Scattering geometry for a prolate spheroid with the confocal
layered structure and $a_1/b_1=2$.
The space is divided into $N+1$ parts:
the outer medium (1), the
outermost layer (2), $\cdots$, the core ($N+1$).
The scattered field in the far-field zone is represented in the spherical
coordinate system $(r,\vartheta,\varphi)$.
$\Theta$ is the scattering angle.
The origin of the Cartesian coordinate system is at the center
of the spheroid while the {\it z} axis coincides with its axis of revolution.
The angle of incidence $\alpha$ is the angle
between the direction of incidence and the {\it z} axis
in the {\it x -- z} plane.

Fig. 2.
Percent difference between three-layered spheres and
three-layered spheroids
$\epsilon$ defined by Eq.~(\ref{eps}):
${m}_3=1.7+0.0i$, ${m}_2=1.5+0.0i$,
${m}_1=1.3+0.0i$, $V_j/V_{\rm total}=0.33$,
$a_{1}/b_{1}=1.0001$, $\alpha=0\degr$,
($\bullet$) -- prolate spheroids,
($\circ$) -- oblate spheroids.

Fig. 3.
Size dependence of the extinction efficiency factors
for layered prolate spheroids with $a_{1}/b_{1}=3$.
Each particle contains an equal fraction
of carbon, silicate, and  vacuum (the porosity  ${\cal P}=1/3$)
separated in equivolume confocal layers.
The cyclic order of the  different material layers is indicated
(starting from the core).
The effect of the increase of the number of layers is illustrated.

Fig. 4.
Size dependence of the normalized extinction cross sections
for 18-layered prolate and oblate spheroids with $a_{1}/b_{1}=3$.
Particles contain an equal fraction
of carbon and silicate without vacuum (the porosity  ${\cal P}=0.0$) or
50\% of vacuum (the porosity  ${\cal P}=0.50$).
For a given value of the size parameter,
the compact and porous particles have the same mass.
The cyclic order of the  different material layers is:
carbon/vacuum/silicate (starting from the core).
The effect of the increase of particle porosity and oblique incidence
is illustrated.

Fig. 5.
The normalized extinction cross sections (see Eq.~(\ref{cn}))
for layered prolate and oblate spheroids with $a_{1}/b_{1}=3$.
For $\alpha=90\degr$, the curves are plotted for the sum of the TM and TE
modes.
The effect of variation of particle type and orientation
is illustrated.

Fig. 6.
The normalized extinction cross sections (see Eq.~(\ref{cn}))
for layered  oblate spheroids.
The effect of variation of particle shape is illustrated.


\newpage
\setcounter{figure}{0}

\begin{figure}\bc
\resizebox{16cm}{!}{\includegraphics{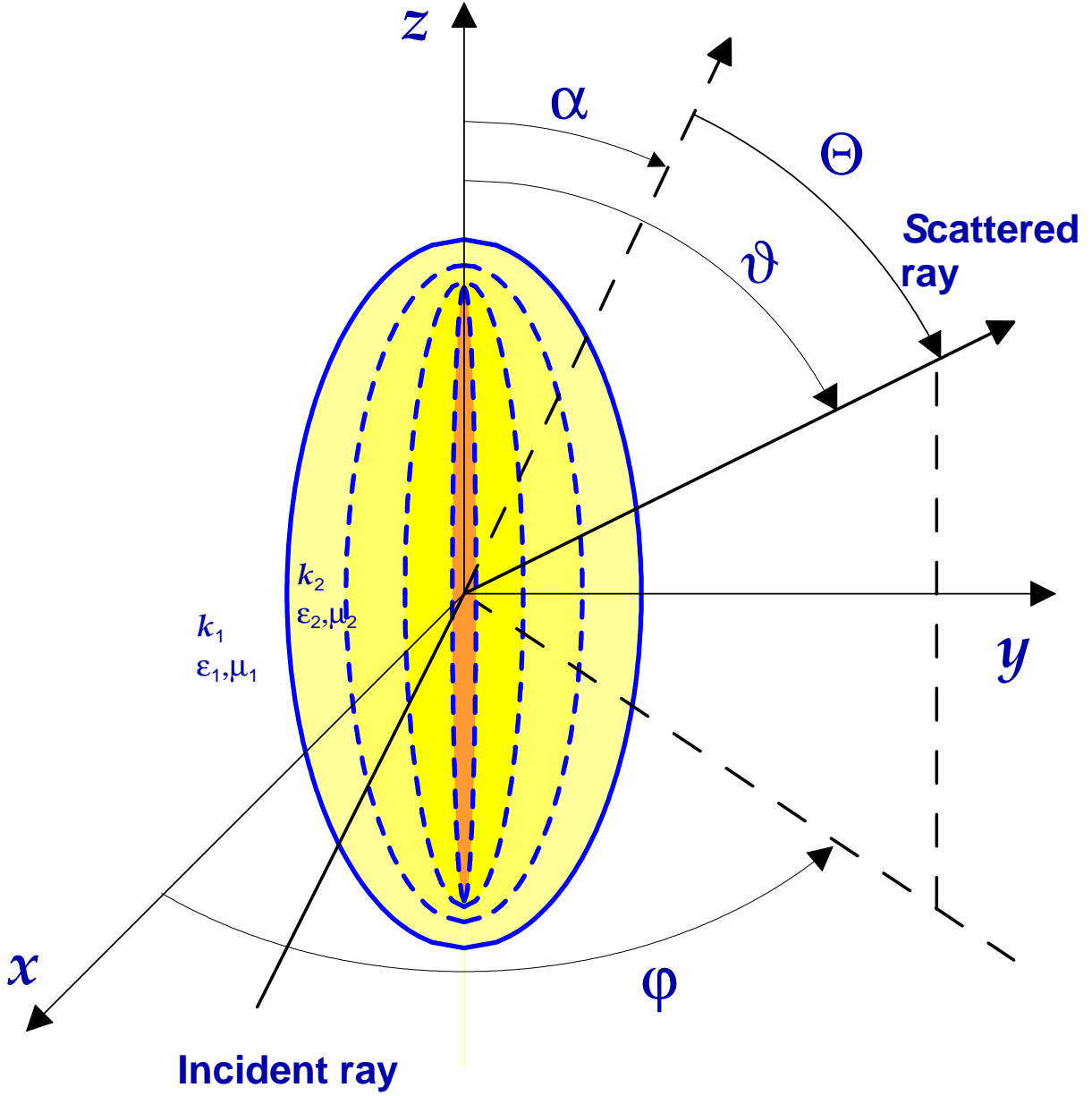}}
\caption{Scattering geometry for a prolate spheroid with the confocal
layered structure and $a_1/b_1=2$.
The space is divided into $N+1$ parts:
the outer medium (1), the
outermost layer (2), $\cdots$, the core ($N+1$).
The scattered field in the far-field zone is represented in the spherical
coordinate system $(r,\vartheta,\varphi)$.
$\Theta$ is the scattering angle.
The origin of the Cartesian coordinate system is at the center
of the spheroid while the {\it z} axis coincides with its axis of revolution.
The angle of incidence $\alpha$ is the angle
between the direction of incidence and the {\it z} axis
in the {\it x -- z} plane.
}\label{geon}
\ec\end{figure}

\begin{figure}\bc
\resizebox{\hsize}{!}{\includegraphics{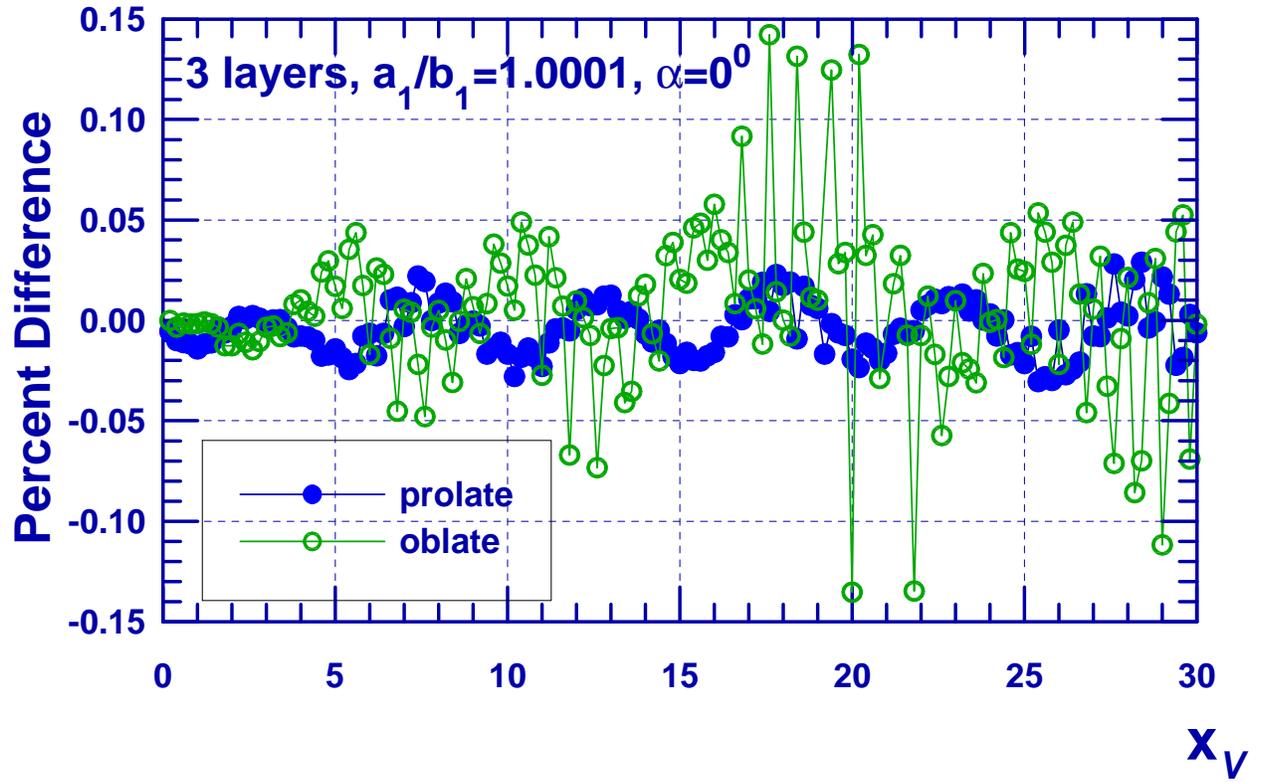}}
\caption{Percent difference between three-layered spheres and
three-layered spheroids
$\epsilon$ defined by Eq.~(\ref{eps}):
${m}_3=1.7+0.0i$, ${m}_2=1.5+0.0i$,
${m}_1=1.3+0.0i$, $V_j/V_{\rm total}=0.33$,
$a_{1}/b_{1}=1.0001$, $\alpha=0\degr$,
($\bullet$) -- prolate spheroids,
($\circ$) -- oblate spheroids.
}\label{ss}
\ec\end{figure}

\begin{figure}\bc
\resizebox{12.0cm}{!}{\includegraphics[bb=48 318 542 546,clip]{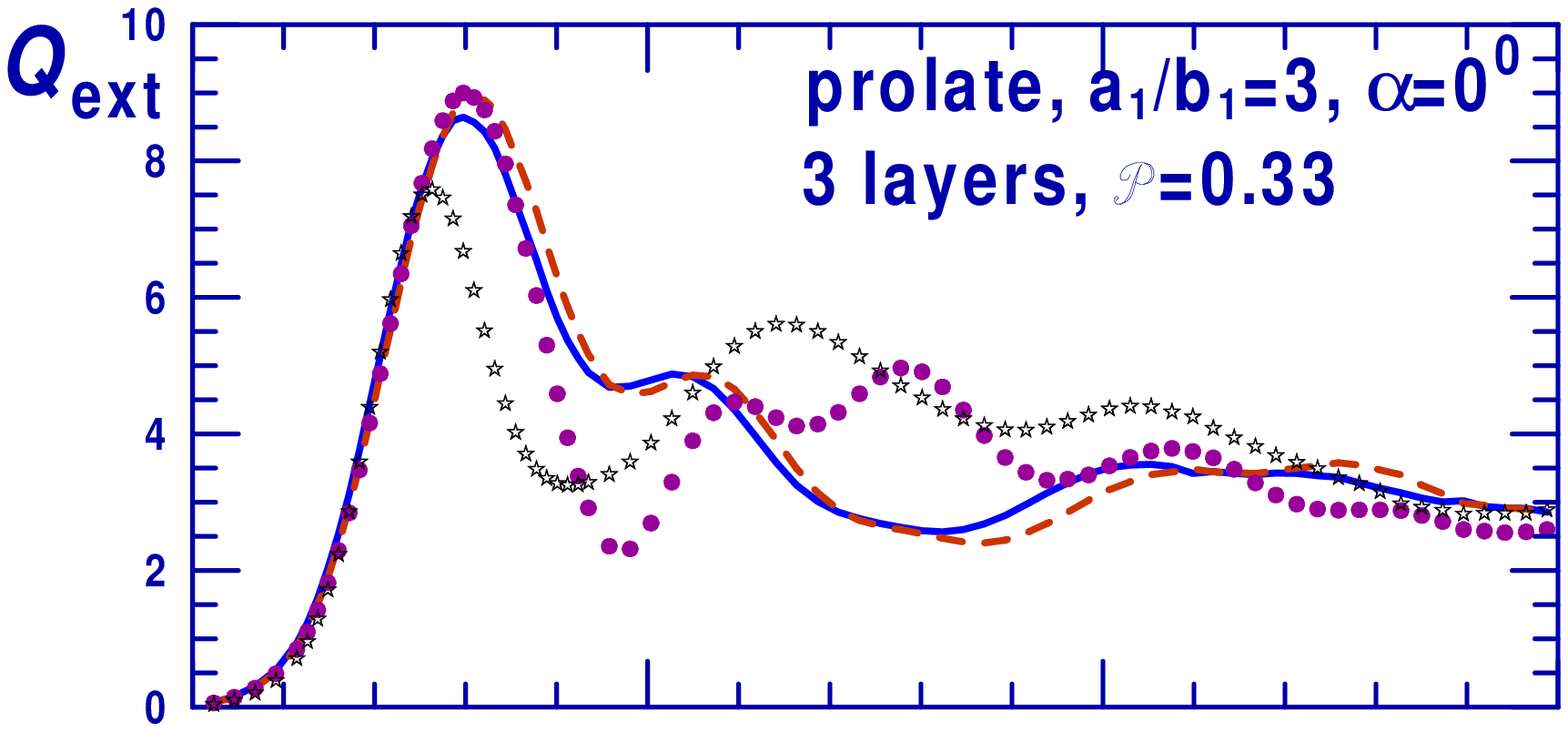}}
\resizebox{12.0cm}{!}{\includegraphics[bb=48 318 542 546,clip]{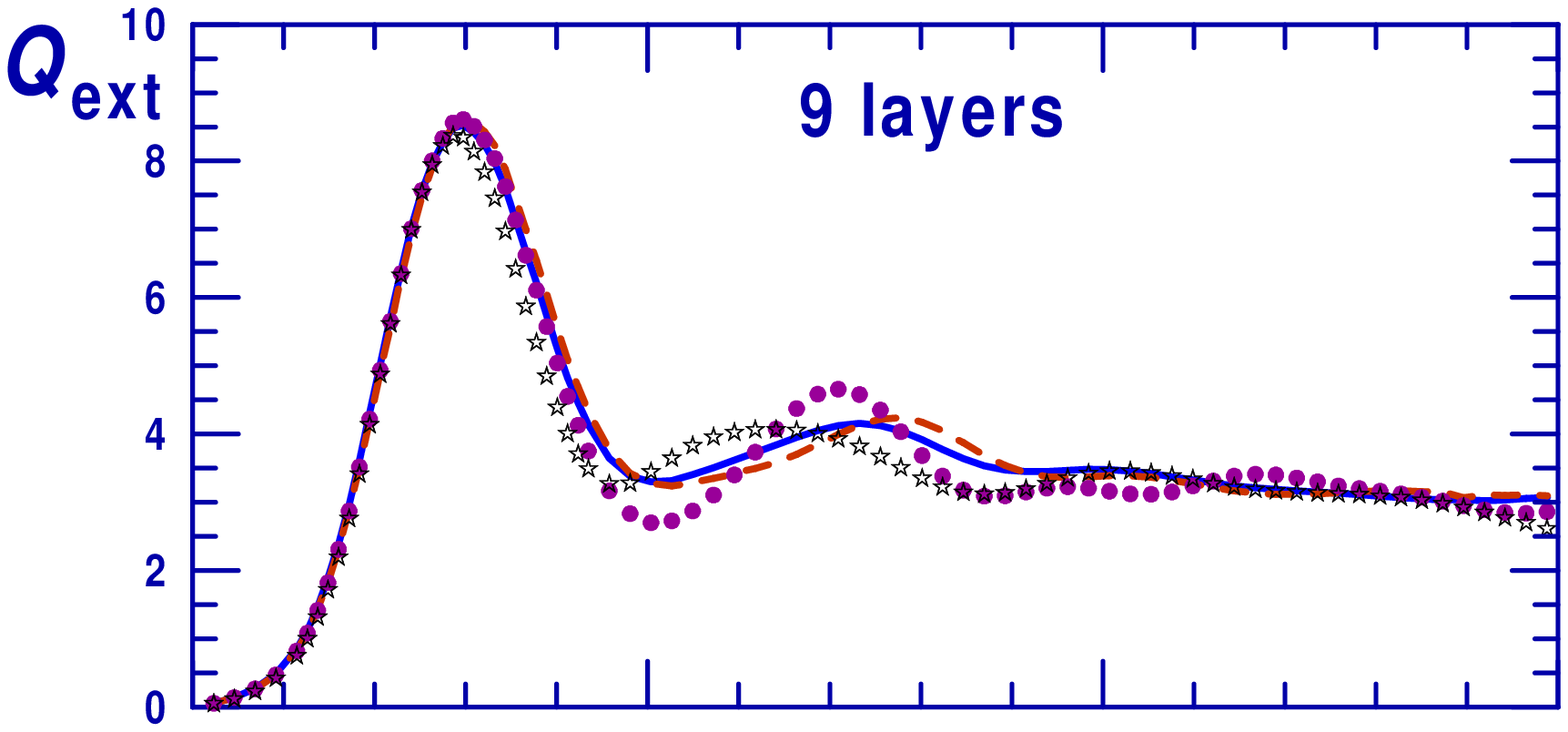}}
\resizebox{12.0cm}{!}{\includegraphics[bb=48 265 545 546,clip]{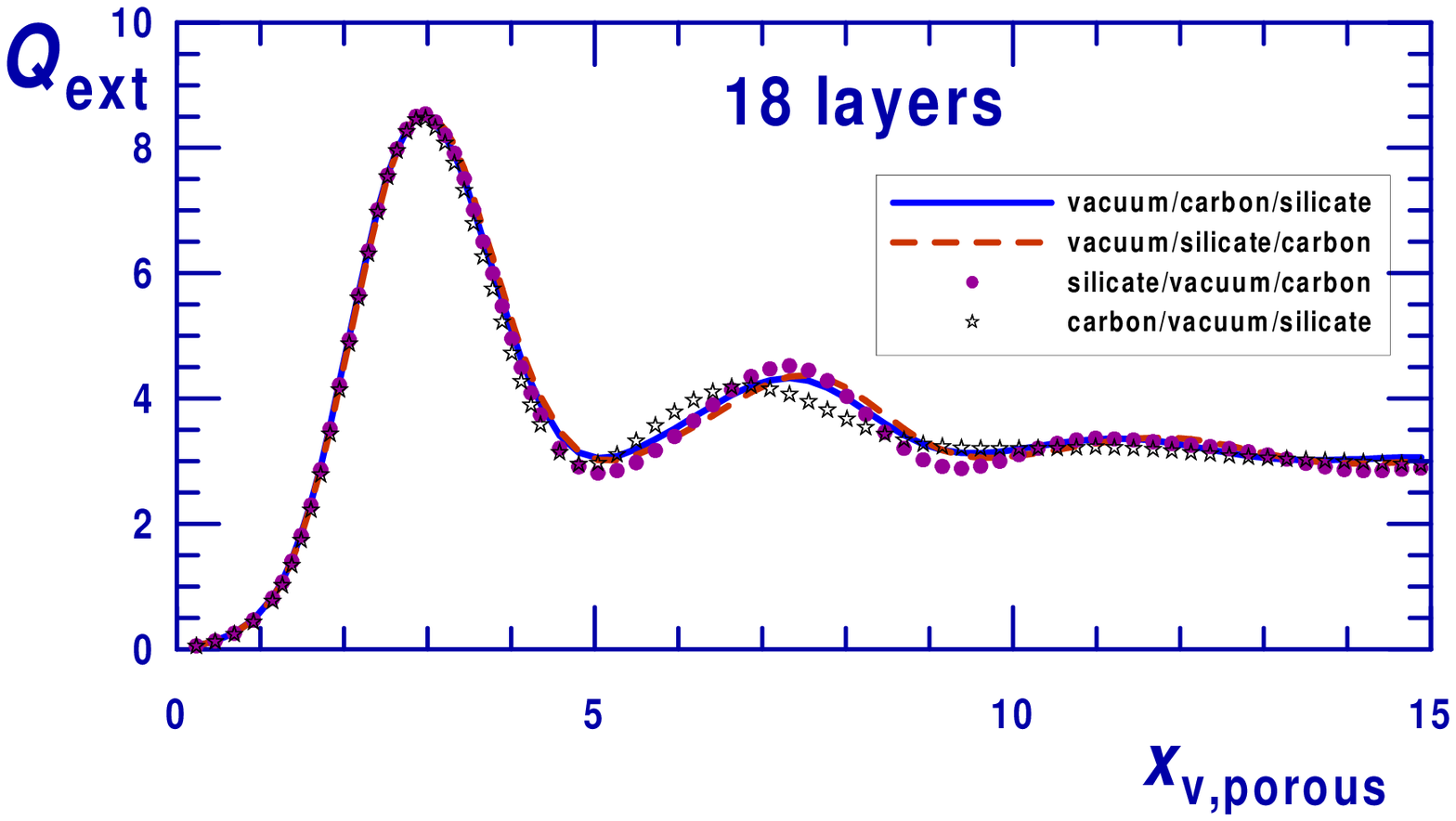}}
\caption{Size dependence of the extinction efficiency factors
for layered prolate spheroids with $a_{1}/b_{1}=3$.
Each particle contains an equal fraction
of carbon, silicate, and  vacuum (the porosity  ${\cal P}=1/3$)
separated in equivolume confocal layers.
The cyclic order of the  different material layers is indicated
(starting from the core).
The effect of the increase of the number of layers is illustrated.
}\label{f3}
\ec\end{figure}

\begin{figure}\bc
\resizebox{12cm}{!}{\includegraphics[bb=58 316 543 545,clip]{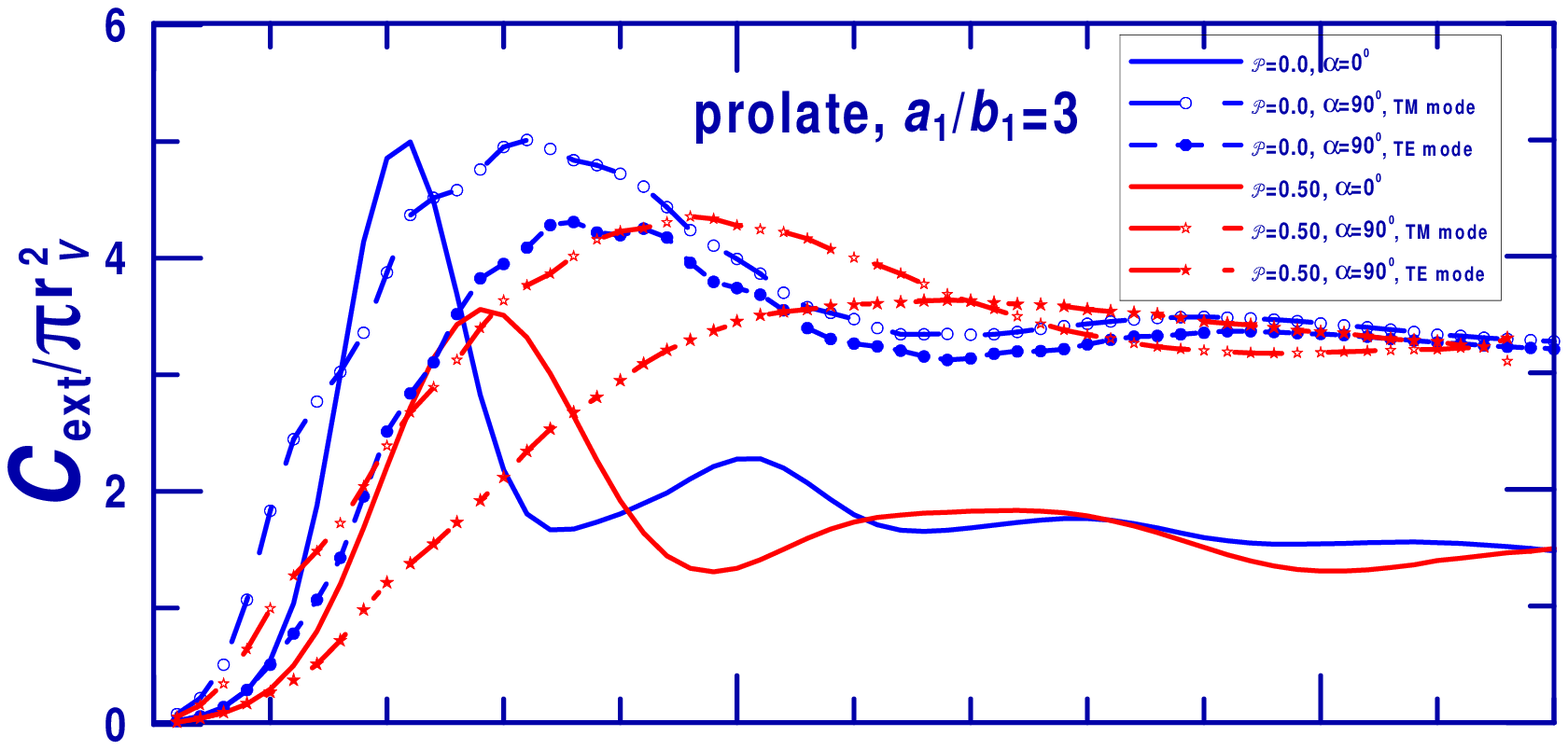}}
\resizebox{12cm}{!}{\includegraphics[bb=58 267 545 547,clip]{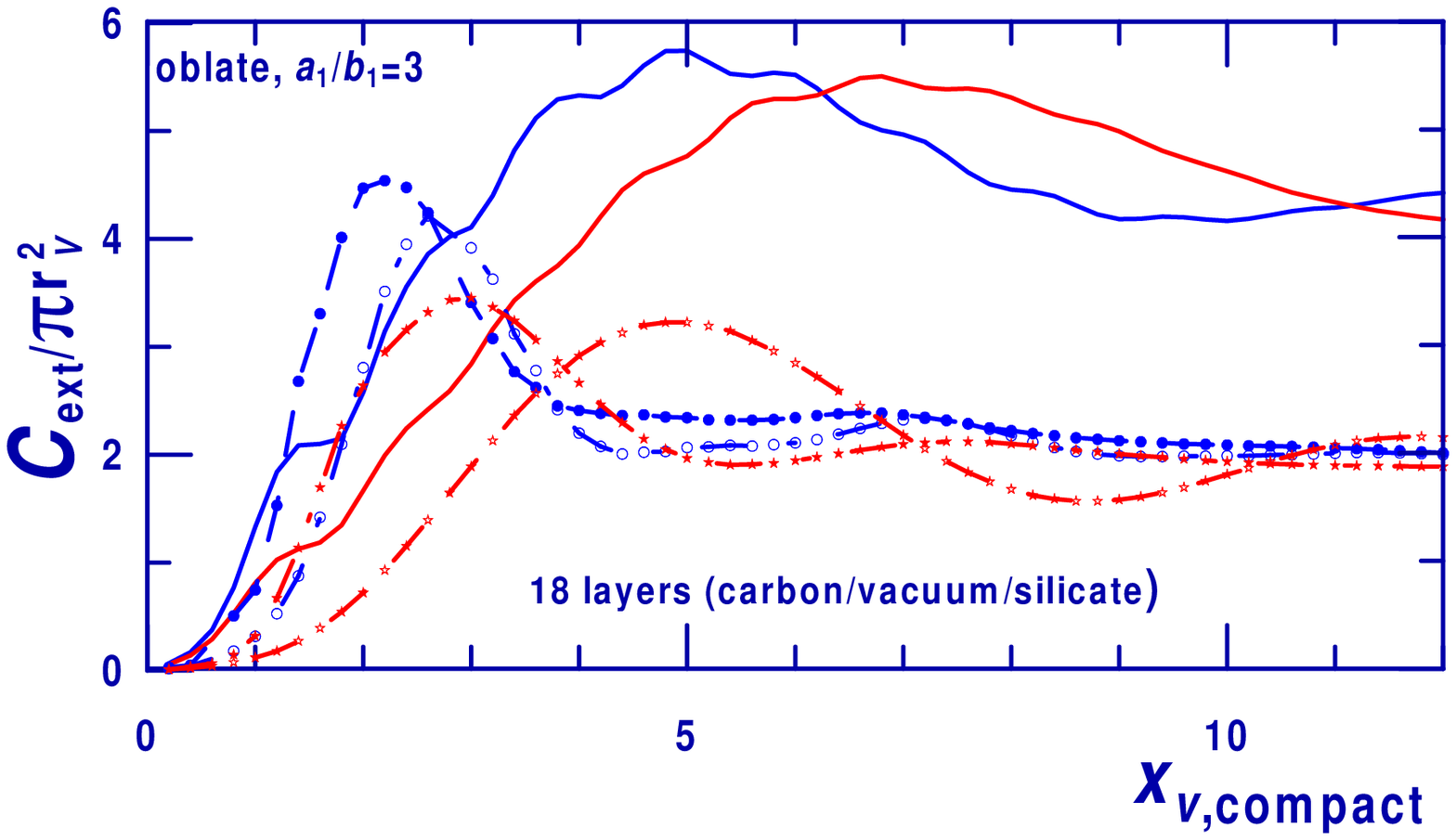}}
\caption{Size dependence of the normalized extinction cross sections
for 18-layered prolate and oblate spheroids with $a_{1}/b_{1}=3$.
Particles contain an equal fraction
of carbon and silicate without vacuum (the porosity  ${\cal P}=0.0$) or
50\% of vacuum (the porosity  ${\cal P}=0.50$).
For a given value of the size parameter,
the compact and porous particles have the same mass.
The cyclic order of the  different material layers is:
carbon/vacuum/silicate (starting from the core).
The effect of the increase of particle porosity and oblique incidence
is illustrated.
}\label{f4}
\ec\end{figure}


\begin{figure}\bc
\resizebox{\hsize}{!}{\includegraphics{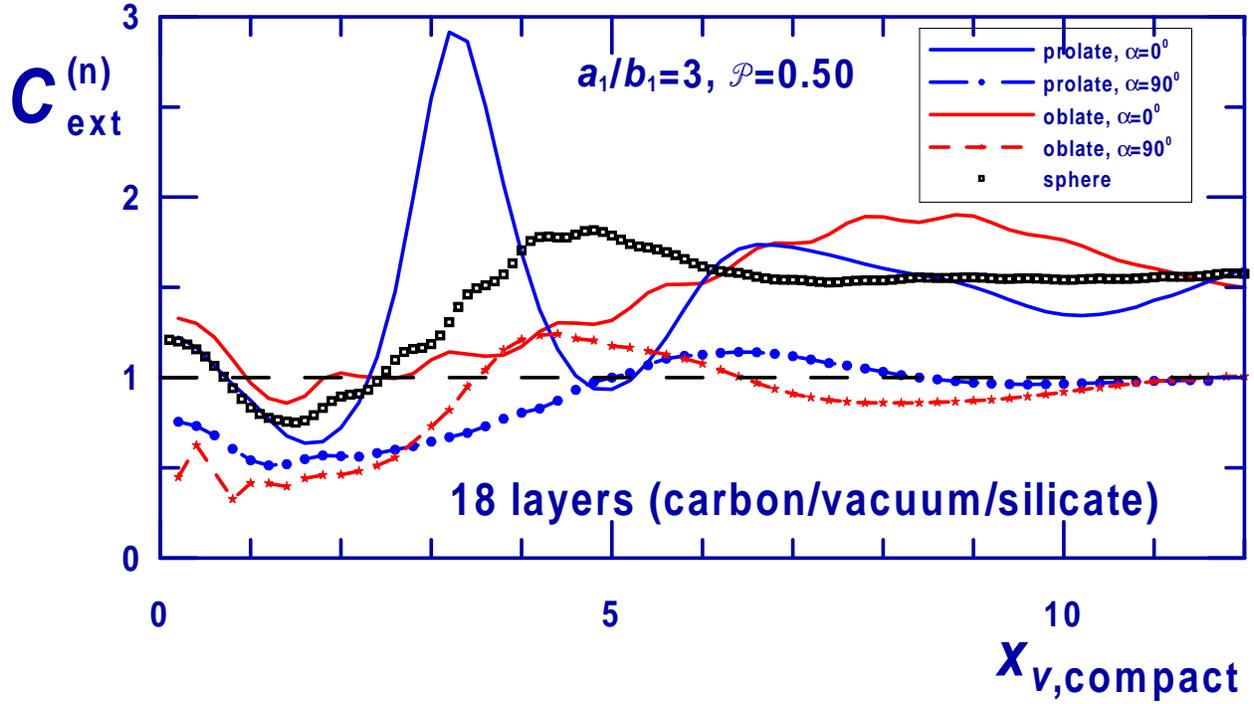}}
\caption{The normalized extinction cross sections (see Eq.~(\ref{cn}))
for layered prolate and oblate spheroids with $a_{1}/b_{1}=3$.
For $\alpha=90\degr$, the curves are plotted for the sum of the TM and TE
modes.
The effect of variation of particle type and orientation
is illustrated.
}\label{f6}
\ec\end{figure}

\begin{figure}\bc
\resizebox{\hsize}{!}{\includegraphics{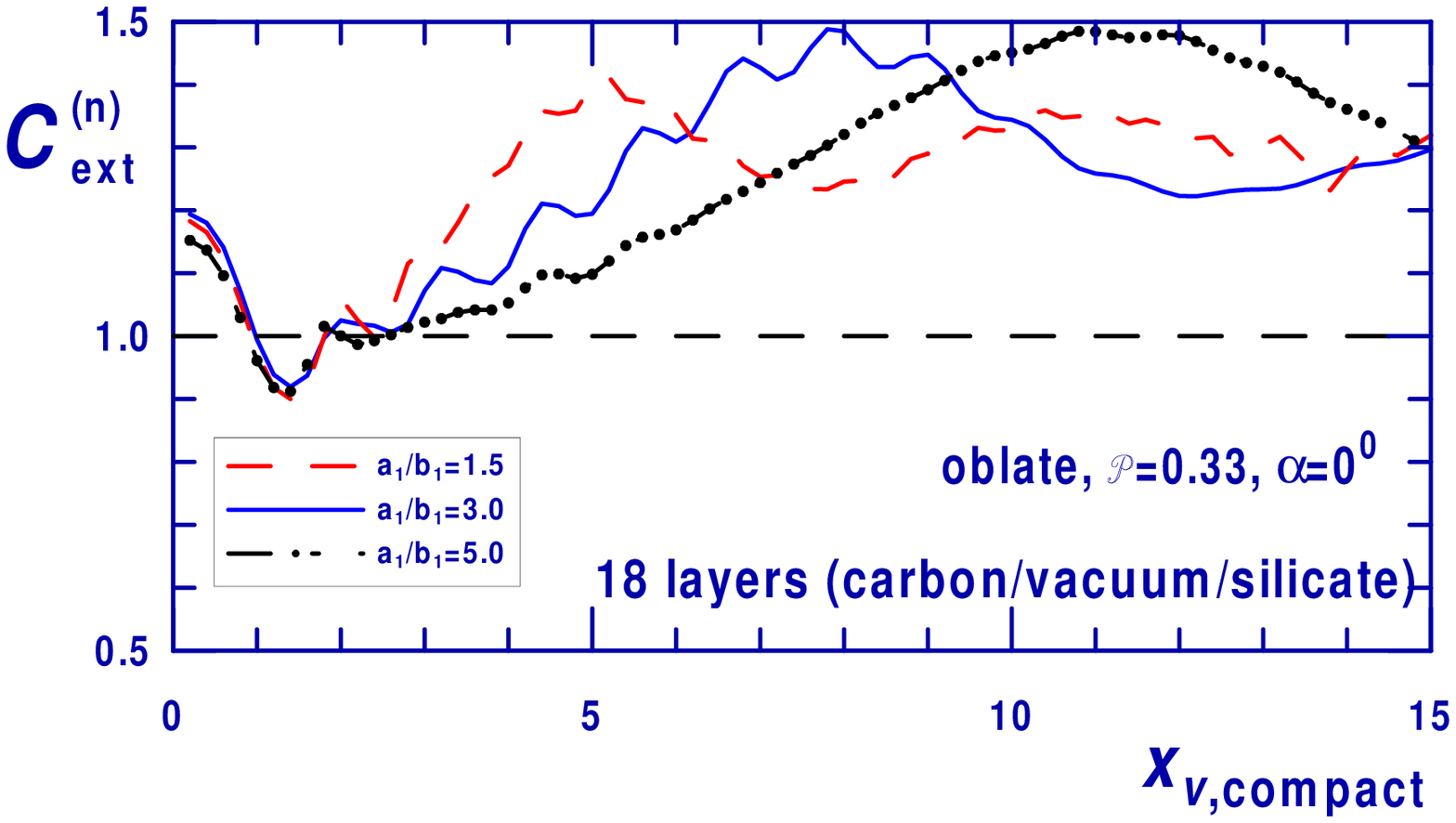}}
\caption{The normalized extinction cross sections (see Eq.~(\ref{cn}))
for layered  oblate spheroids.
The effect of variation of particle shape is illustrated.
}\label{f7}
\ec\end{figure}


\begin{references}


\bibitem{fip03}  V. G. Farafonov, V. B. Il'in, and M. S. Prokopjeva,
  ``Light scattering by multilayered nonspherical particles: a set
  of methods,'' \jqsrt \textbf{79--80},
  599--626 (2003).

\bibitem{kan03}  F. M. Kahnert,
  ``Numerical methods in electromagnetic scattering theory,''
  \jqsrt \textbf{79--80}, 775--824 (2003).

\bibitem{vfi09}  A. Vinokurov, V. Farafonov,  and V. Il'in,
  ``Separation of variables method for multilayered nonspherical particles,''
  \jqsrt \textbf{110},   1356--1368 (2009).

\bibitem{f11}   V. G. Farafonov,
  ``A unified approach, using spheroidal  functions, for solving the
problem of light scattering by a axisymmetric particles,''
J. Math. Sci. \textbf{175},   698--723 (2011).


%
%
%
%
%

\bibitem{getal00}
I. Gurwich, M. Kleiman, N. Shiloah, and A. Cohen, ``Scattering of
electromagnetic radiation by multilayered spheroidal particles:
recursive procedure,''
\ao \textbf{39}, 470--477 (2000).

\bibitem{far01}
V. G. Farafonov, ``New recursive solution to the problem of scattering
of electromagnetic radiation by multilayered spheroidal particles,''
Opt. Spectrosc.  \textbf{90}, 743--752 (2001).


\bibitem{getal03}
I. Gurwich, M. Kleiman, N. Shiloah, and D. Oaknin, ``Scattering by
an arbitrary multi-layered spheroid: theory and numerical results,''
\jqsrt \textbf{79--80}, 649--660 (2003).

\bibitem{han06}
Y. Han, H. Zhang, and X. Sun, ``Scattering of
shaped beam by an arbitrarily oriented
spheroid having layers with non-confocal boundaries,''
\ap \textbf{B 84}, 485--492 (2006).

\bibitem{wb79}
D. S. Wang, P. W. Barber,
``Scattering by inhomogeneous nonspherical objects'',
\ao \textbf{18}, 1190--1197 (1979).

\bibitem{pet07}
D. Petrov, Y. Shkuratov, E. Zubko, and G. Videen,
``Sh-matrices method as applied to scattering
by particles with layered structure'',
\jqsrt \textbf{106}, 437--454 (2007).

\bibitem{dwe06}
A. Doicu, T. Wriedt, and Y. Eremin,
{\em Light scattering by systems of particles}
(Springer, 2006).

\bibitem{nvv06}
N. V. Voshchinnikov, V. G. Farafonov, G. Videen, and L. S. Ivlev,
``Development of the separation of variables method for multi-layered
spheroids'', in {\em Proc. of the 9th Conf. on
Electromagnetic and Light Scattering by Nonspherical Particles},
N. V. Voshchinnikov, ed. (St. Petersburg Univ., 2006), pp.~271--274.

\bibitem{fvs96}
V. G. Farafonov, N. V. Voshchinnikov, and V. V.  Somsikov,
``Light scattering by a core-mantle spheroidal particle,''
\ao \textbf{35}, 5412--5426 (1996).


\bibitem{F57}
C. Flammer, {\em Spheroidal  Wave  Functions}
(Stanford  Univ.   Press, 1957).

\bibitem{KPS76}
I. V. Komarov, L. I. Ponomarev, and S. Yu. Slavyanov,
   {\em Spheroidal and Coulomb Spheroidal Functions}
   (Nauka, 1976).

\bibitem{fi06} V. G. Farafonov and V. B. Il'in,
``Single light scattering: computational methods'',
in {\em Light Scattering Reviews},  A. A. Kokhanovsky, ed.
(Springer, 2006) \textbf{1}, pp. 125--177.

\bibitem{vf93} N. V. Voshchinnikov and  V. G. Farafonov,
``Optical  properties of spheroidal particles'',
Astrophys. Space Sci. \textbf{204}, 19--86 (1993).

\bibitem{vf02}  N. V. Voshchinnikov and V. G. Farafonov,
``Numerical treatment of spheroidal wave functions,''
 In: \textit{Electromagnetic and Light Scattering by Nonspherical Particles},
B. \AA. S. Gustafson  et al. (eds.),
 Army Res. Lab., Adelphi, 325--328 (2002).

\bibitem{vf04}  N. V. Voshchinnikov and V. G. Farafonov,
``Computation of radial prolate spheroidal wave functions
 using J\'affe's series expansions'',
J.  Comp. Math. Math. Phys. \textbf{43}, 1299--1309 (2003).

\bibitem{vm99} N. V. Voshchinnikov and J. S. Mathis,
``Calculating cross sections of composite interstellar grains,''
\apj \textbf{526}, 257--264 (1999).

\bibitem{Po_02} B. Posselt,  V. G. Farafonov, V. B. Il'in, and M. S. Prokopjeva,
  ``Light scattering by multilayered ellipsoidal particles
  in the quasistatic approximation'',
  Measurements and Sci. Technol. \textbf{13}, 256--262 (2002).

\bibitem{vih04} N. V. Voshchinnikov, V. B. Il'in, and Th. Henning,
``Modelling the optical properties of composite and porous interstellar grains'',
  Astronomy and Astrophysics \textbf{429}, 371--381  (2005).

\bibitem{ks94} E. Kr\"ugel and R. Siebenmorgen,
``Dust in protostellar cores and stellar disks'',
  Astronomy and Astrophysics \textbf{288}, 929--941  (1994).



\end{references}
\end{document}